\documentclass[letterpaper,english,reprint, aps,longbibliography,superscriptaddress]{revtex4-1}
\usepackage[T1]{fontenc}
\usepackage[latin9]{inputenc}
\usepackage{color}
\usepackage{babel}
\usepackage{amsmath}
\usepackage{mathtools}
\usepackage{amssymb}
\usepackage[mathscr]{euscript}
\usepackage[unicode=true,
 bookmarks=true,bookmarksnumbered=false,bookmarksopen=false,
 breaklinks=false,pdfborder={0 0 1},backref=false,colorlinks=false]
 {hyperref}

\usepackage{bbold}

\makeatletter

\pdfpageheight\paperheight
\pdfpagewidth\paperwidth

\makeatletter
\newcommand{\oalpha}{\mathbin{\mathpalette\make@circled{{\alpha}}}}
\newcommand{\make@circled}[2]{  \ooalign{$\m@th#1\smallbigcirc{#1}$\cr\hidewidth$\m@th#1#2$\hidewidth\cr}}
\newcommand{\smallbigcirc}[1]{  \vcenter{\hbox{\scalebox{1.5}{$\m@th#1\bigcirc$}}}}
\makeatother

\makeatletter
\usepackage{tikz}
\newcommand*\circled[2][1.6]{\tikz[baseline=(char.base)]{
    \node[shape=circle, draw, inner sep=1pt, 
        minimum height={\f@size*#1},] (char) {#2};}}
\makeatother

\newcommand{\kT}{k_{\mathrm{B}}T}

\newcommand{\TimeDiff}{\Delta t_{\mathrm{tot}}}

\newcommand{\traj}{\varphi}

\newcommand{\trajTwo}{\psi}

\newcommand{\PR}{P_{R}^{\traj}}
\newcommand{\dotPR}{\dot{P}_{R}^{\traj}}
\newcommand{\PRTwo}{P_{R}^{\trajTwo}}

\newcommand{\tinitial}{t_i}
\newcommand{\tfinal}{t_f}

\newcommand{\dparam}{\xi}

\newcommand{\aexit}{\alpha_{R}}

\newcommand{\SPI}{\mathscr{S}}
\newcommand{\Bb}{\mathcal{B}}

\newcommand{\xminleft}{x^{\mathrm{min}}_{0}}
\newcommand{\xminright}{x^{\mathrm{min}}_{1}}

\newcommand{\RefillTime}{\Delta \mathcal{T}}

\makeatother

\begin{document}
\title{ Experimental measurement of relative path probabilities and  stochastic actions}
\author{Jannes Gladrow}
\affiliation{Microsoft Research, Station Rd, Cambridge CB1 2FB, United Kingdom}
\affiliation{Cavendish Laboratory, University of Cambridge, JJ Thomson Ave, Cambridge CB3 0HE, United Kingdom}
\author{Ulrich F.~Keyser}
\affiliation{Cavendish Laboratory, University of Cambridge, JJ Thomson Ave, Cambridge CB3 0HE, United Kingdom}
\author{R.~Adhikari}
\author{Julian Kappler}
\email{jkappler@posteo.de}
\affiliation{DAMTP, Centre for Mathematical Sciences, University of Cambridge, Wilberforce Road, Cambridge CB3 0WA, United Kingdom}

\date{\today}
\begin{abstract}
For diffusive stochastic dynamics, the probability to observe any individual trajectory is vanishingly small, making it unclear how to experimentally validate theoretical results for ratios of path probabilities.
We provide the missing link between theory and experiment, by establishing 
a protocol to extract ratios of path probabilities from measured time series.
For experiments on a single colloidal particle in a microchannel,
we extract both ratios of path probabilities, and the most probable path for a barrier crossing\textcolor{black}{,
and find} excellent agreement with independently calculated predictions based on
the Onsager-Machlup stochastic action.
\textcolor{black}{Our experimental results at room temperature
 are found to be inconsistent with the low-noise Freidlin-Wentzell stochastic action,
 and we discuss under which circumstances the latter is expected to describe
 the most probable path.}
 \textcolor{black}{
Furthermore, while the experimentally accessible ratio of path probabilities is uniquely determined,
 the formal path-integral action is known to depend on the time-discretization scheme used
for deriving it;
  we reconcile these two seemingly contradictory facts by careful
 analysis of the time-slicing derivation of the path integral.}
 \textcolor{black}{Our experimental protocol
enables us to 
 probe probability distributions on path space,
 and 
allows us to relate
 theoretical single-trajectory results to measurement.}
\end{abstract}
\maketitle

\section{Introduction}
Stochastic effects are of fundamental relevance for statistical physics
and beyond \cite{gardiner_stochastic_2009,oksendal_stochastic_2007,kampen_stochastic_2007,
sekimoto_stochastic_2010,dembo_large_2010,chetrite_nonequilibrium_2015,nolting_balls_2016,
bruckner_stochastic_2019,friz_large_2015}.
For example, diffusion processes are used to model colloidal particles \cite{sekimoto_stochastic_2010,
seifert_stochastic_2012,
bera_fast_2017}, 
polymer dynamics \cite{wilemski_diffusioncontrolled_1974-1,
szabo_first_1980,
doi_theory_2007,
vandebroek_generalized_2017},
or active particles such as driven colloidal systems, cells, or bacteria
\cite{seifert_stochastic_2012,
bruckner_stochastic_2019}.
Any stochastic dynamics is fully characterized by its path probabilities,
which are also highly relevant in applications;
examples are irreversibility in stochastic thermodynamics, which is expressed
in terms of ratios of path probabilities \cite{seifert_entropy_2005,seifert_stochastic_2012},
or transition pathways between metastable states \cite{luchinsky_analogue_1998,
chan_paths_2008}, as relevant e.g.~for conformational
transitions in biomolecules \cite{e_string_2002,ren_transition_2005,e_transition_2005}.

 For diffusive dynamics, the probability \textcolor{black}{to observe any given individual}
path is zero. Still, ratios of path probabilities can be quantified
theoretically by stochastic actions \cite{onsager_fluctuations_1953,stratonovich_probability_1971,durr_onsager-machlup_1978,fujita_onsager-machlup_1982,horsthemke_onsager-machlup_1975,ito_probabilistic_1978,williams_probability_1981,ikeda_stochastic_1989,graham_path_1977,langouche_functional_1979,dekker_functional_1978,
dekker_path_1980,weber_master_2017,wissel_manifolds_1979,adib_stochastic_2008,
cugliandolo_building_2019}.
The literature contains
several proposals for stochastic actions, the prominent ones being
associated with the names of Onsager and Machlup (OM) \cite{onsager_fluctuations_1953,
stratonovich_probability_1971,durr_onsager-machlup_1978,fujita_onsager-machlup_1982},
as well as Freidlin and Wentzell (FW) \cite{ventsel_small_1970,touchette_large_2009,grafke_numerical_2019}.
 Since it is not straightforward to 
\textcolor{black}{access probability-zero events
in measurement,
hitherto it was not clear how to directly
validate theoretical results for stochastic actions.}

We here overcome this difficulty, by
establishing an experimental protocol to determine ratios of path probabilities from 
observed data, without fitting a model \textcolor{black}{to the stochastic dynamics}.
We achieve this by considering the
 sojourn probability
\cite{stratonovich_probability_1971,fujita_onsager-machlup_1982,ventsel_small_1970},
i.e.~the
  probability
 that a stochastic trajectory remains within a tube of small but finite radius $R$
 around a reference path \textcolor{black}{[see Fig.~\ref{fig:extrapolated_rate} (a) for an illustration]}.
For a colloidal particle in a microchannel subject to a double-well potential,
we directly measure the finite-radius sojourn probability for a pair of reference paths,
and subsequently extrapolate the ratio of sojourn probabilities to the limit $R \rightarrow 0$.
We demonstrate that this experimentally observed ratio of path probabilities is well-described 
by the difference in OM Lagrangians along the two reference paths,
thereby 
\textcolor{black}{confirming classical theoretical results on the asymptotic sojourn
probability \cite{stratonovich_probability_1971,durr_onsager-machlup_1978,
fujita_onsager-machlup_1982} and}
transforming the OM action from a purely mathematical construct into a physical observable.
We observe that our results for relative path probabilities \textcolor{black}{at room temperature}
are markedly different from
the predictions of the FW Lagrangian \cite{ventsel_small_1970,touchette_large_2009,grafke_numerical_2019},
\textcolor{black}{which 
in the context of sojourn probabilities is associated with a low-noise limit \cite{ventsel_small_1970}.}
Considering the most probable path, or instanton, as
zero-radius limit of the most probable tube, 
we furthermore determine the instanton
for experimental barrier-crossing paths in a double-well potential, which is again well-described by
the OM Lagrangian, and different from the FW prediction.
We discuss quantitatively for which system parameters the FW Lagrangian \cite{ventsel_small_1970}
is expected
to describe the physical most probable path.
\textcolor{black}{Finally, 
we resolve the  
 seeming contradiction that the asymptotic-tube Lagrangian is uniquely defined,
 whereas the formal stochastic action Lagrangian that appears in the path-integral formalism is not.}
\textcolor{black}{
Our results demonstrate that
ratios of path probabilities can be inferred from experimental data
without the need to fit a stochastic model.
}

\section{Theory and experimental results}

\subsection{Sojourn probability and stochastic action}
\textcolor{black}{
We consider the approach to relative path probabilities via  the sojourn probability $\PR(t)$, 
i.e.~the probability
 that a stochastic
trajectory $X_t \equiv X(t)$ remains within a moving ball, of radius $R$ and 
with a center parametrized by a twice-differentiable reference path
$\traj_t \equiv \traj(t)$, up to a time $t$ \cite{ventsel_small_1970,stratonovich_probability_1971,
horsthemke_onsager-machlup_1975,
durr_onsager-machlup_1978,ito_probabilistic_1978,
williams_probability_1981,fujita_onsager-machlup_1982,ikeda_stochastic_1989}.
}
The relative likelihood   for two given reference paths $\traj(t)$, $\trajTwo(t)$, $t \in [t_i,t_f]$,
\textcolor{black}{is then} quantified by a stochastic action $S$, defined via \cite{ventsel_small_1970,stratonovich_probability_1971,
horsthemke_onsager-machlup_1975,
durr_onsager-machlup_1978,ito_probabilistic_1978,
williams_probability_1981,fujita_onsager-machlup_1982,ikeda_stochastic_1989}
\begin{equation}
	\label{eq:ActionDifference}
	\frac{e^ {-S[\traj]} }{e^ {-S[{\trajTwo}]}} \equiv \lim_{R \rightarrow 0}  \frac{\PR(\tfinal)}{\PRTwo(\tfinal)}.
\end{equation}
For Markovian dynamics, the action is the integral 
over a Lagrangian \cite{fujita_onsager-machlup_1982,ikeda_stochastic_1989},
\begin{equation}
\label{eq:LagrangianDef}
	S[{\traj}] = \int_{\tinitial}^{\tfinal} \mathrm{d}t~\mathcal{L}^{\traj}(t),
\end{equation}
 \textcolor{black}{
and for  overdamped Langevin dynamics with additive noise,
Stratonovich was the first to show  that 
 the limiting ratio of sojourn probabilities 
is described by the OM
 Lagrangian \cite{stratonovich_probability_1971};
this analytical result has since been confirmed by several
 subsequent derivations \cite{durr_onsager-machlup_1978,
ito_probabilistic_1978,
williams_probability_1981,
fujita_onsager-machlup_1982,ikeda_stochastic_1989,
kappler_stochastic_2020}.
A notable exception is the derivation of Freidlin and Wentzell \cite{ventsel_small_1970}, 
where the FW Lagrangian is obtained as
the sojourn probability in a low-noise limit.
}

\textcolor{black}{
A different route to quantifying path probabilities via stochastic actions is given by
the path-integral formalism, 
in which the propagator of a stochastic dynamics
is formally represented as an integral over all paths connecting an initial
and a final point \cite{onsager_fluctuations_1953,
graham_path_1977,
langouche_functional_1979,
dekker_path_1980,
weber_master_2017,
wissel_manifolds_1979,
adib_stochastic_2008,
cugliandolo_building_2019}.
In the standard time-slicing derivation of the path-integral formalism,
a formal stochastic action is defined via the continuum limit of concatenated
short-time propagators.
The formal limit is not uniquely defined, but even for overdamped Langevin dynamics
with constant diffusivity (additive noise)
depends on the discretization used for the short-time propagators;
this leads to the FW and OM actions as two of infinitely many equivalent representations
of the path-integral action
\cite{haken_generalized_1976,
wissel_manifolds_1979,
langouche_functional_1979,
adib_stochastic_2008}.
While this non-uniqueness 
seems in conflict with the definite finite-temperature limit found in
Eq.~\eqref{eq:LagrangianDef}, we will, in Sect.~\ref{sec:path_integral_action} further below, 
 resolve this apparent contradiction
by careful analysis of the time-slicing path integral derivation.
}

\textcolor{black}{
The practical relevance of the asymptotic sojourn probability Eq.~\eqref{eq:ActionDifference}
is that it directly leads to a simple relation
between the stochastic action and physical observables.}
To make this explicit,
we substitute Eq.~\eqref{eq:LagrangianDef} into the logarithm of Eq.~\eqref{eq:ActionDifference}, 
differentiate the result 
with respect to $\tfinal$, and subsequently rename $\tfinal$ to $t$.
This yields
\begin{equation}
\label{eq:LagrangianDifference}
\mathcal{L}^{\traj}(t) - \mathcal{L}^{\trajTwo}(t)
=
 \lim_{R \rightarrow 0} \left(\vphantom{\frac{1}{2}} \aexit^{\traj}(t) - \aexit^{\trajTwo}(t) \right),
\end{equation}
where the instantaneous exit rate at which stochastic trajectories  first
leave the ball of radius $R$ around $\traj$ is given by
	  $\aexit^{\traj}(t)\equiv - \dotPR(t)/\PR(t)$,
where a dot denotes a derivative with respect to time $t$.
For a finite radius $R$ the right-hand side of Eq.~\eqref{eq:LagrangianDifference}
 can be measured directly experimentally,
  via the ratio of recorded trajectories which remain within the threshold
 distance $R$ to $\traj$ as a function of time.
Thus, 
Eq.~\eqref{eq:LagrangianDifference}
 yields a \textcolor{black}{model-free} 
 experimental route to action Lagrangian differences,
 via extrapolating  exit-rate differences measured at finite radius
 to the limit $R \rightarrow 0$.
\textcolor{black}{As we show in the following,
this allows to measure stochastic Lagrangian differences,
and hence to experimentally
test theoretical predictions for ratios of path probabilities.
}

\subsection{Experimental setup}
\begin{figure}[ht!]
\centering
	\includegraphics[width=0.7\columnwidth]{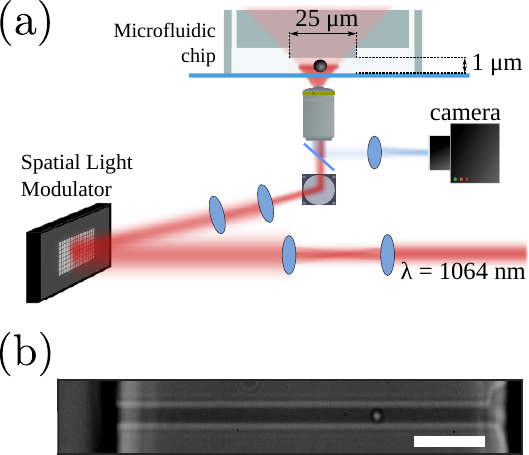}\\
	\vspace{0.5cm}
	\includegraphics[width=0.73\columnwidth]{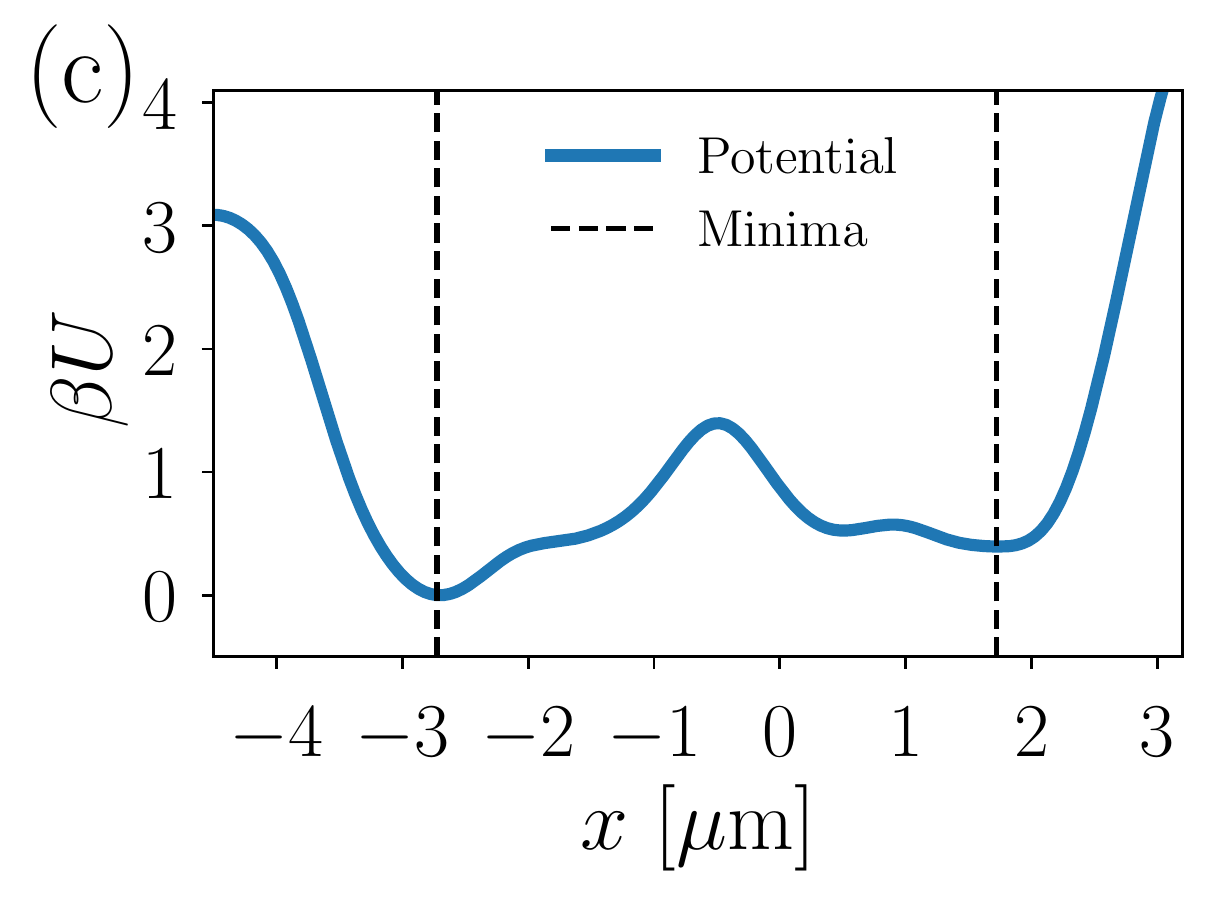}
\caption{
\label{fig:setup_and_potential}
\textit{(a) Experimental setup.}
In our experiments, following the general setup of Ref.~\cite{chupeau_optimizing_2020},
 we observe the motion of a colloidal particle inside a microchannel.
\textit{(b) Image of colloidal particle in microchannel.}
The motion of the colloidal particle can be considered 
effectively one-dimensional.
The horizontal scale bar in the lower right corner is $5\,\mu$m in length,
the colloidal particle has a diameter of $500$ nm. 
Subfigures (a) and (b) are adapted from Ref.~\cite{chupeau_optimizing_2020}.
\textit{(c) Potential energy extracted from experimental time series.}
The blue solid line depicts the potential energy, obtained from 
evaluating the first two Kramer-Moyal 
coefficients based on experimental data, and subsequent smoothing as explained in 
Appendix \ref{app:parametrization}. 
The vertical dashed lines denote two local minima of the potential energy landscape,
located at $\xminleft \approx -2.7\,\mu$m, $\xminright \approx 1.7\,\mu$m.
}
\end{figure}

In our experiments, we observe the motion
of a colloidal particle confined to a microchannel.
Our experimental setup, illustrated in Fig.~\ref{fig:setup_and_potential}, 
consists of a holographic optical tweezer, which 
can autonomously capture colloidal particles and position them inside a microchannel
filled with aqueous salt solution.
Due to the strong confinement created by the channel, 
the motion of the center point of the colloidal particle can be considered 
effectively one-dimensional, i.e.~along the channel axis.
A spatial light modulator is used to form an optical landscape 
which gives rise to a potential energy landscape
that 
the colloid experiences inside the channel.
We tune the modulator to create an approximate double-well potential, shown in
Fig.~\ref{fig:setup_and_potential} (c);
\textcolor{black}{on the scales we probe, this potential is time-independent}.
The position of the colloidal particle is recorded 
at 1000 frames per second. In total, we record and analyze
approximately 104 minutes of experimental measurements, partitioned into short trajectories of variable 
length ranging from 10 to 60 seconds.
The experimental setup is discussed in more detail in Refs.~\cite{gladrow_experimental_2019,chupeau_optimizing_2020}.

\subsection{Relative path likelihoods from experiment}
\begin{figure}[ht!]
\centering
	\includegraphics[width=\columnwidth]{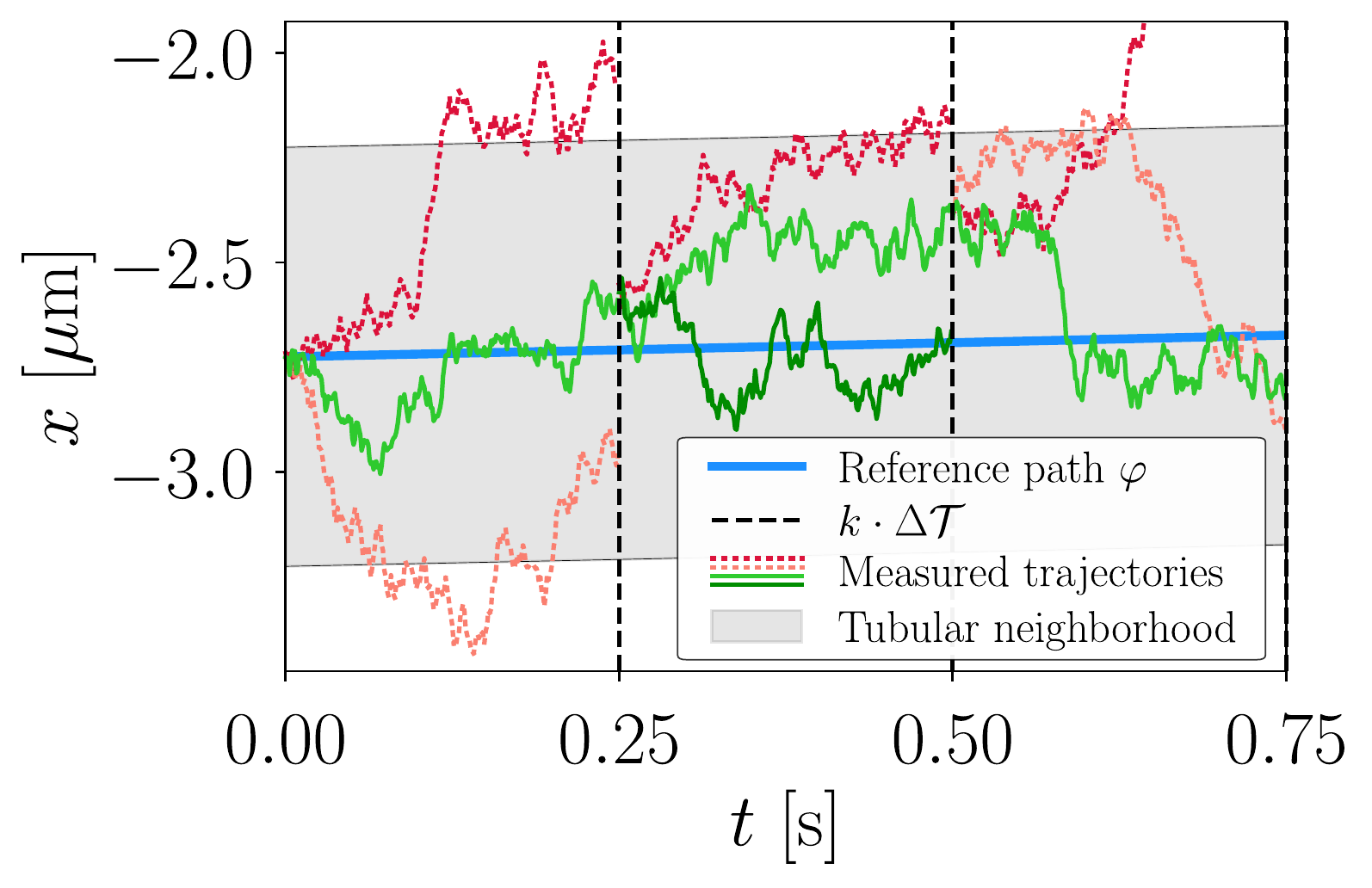}
\caption{
\label{fig:illustration}
\textit{Illustration of our algorithm for obtaining sojourn probabilities from measured
 time series.}
The solid blue line represents a reference path $\varphi$, 
around which a tube of radius $R = 0.5\,\mu$m is shown as grey shaded area.
We randomly select $M=3$ measured trajectories which start in
a small interval around $\xminleft \approx -2.7\,\mu\mathrm{m}$,
and follow them for a duration $\RefillTime = 0.25\,\mathrm{s}$ (vertical dashed lines).
Trajectories which leave the tube
(red dotted lines) are discarded,
the final positions of those trajectories that stay (green solid lines)
are collected.
We then again randomly select $M=3$ measured trajectories, which start in a 
small interval around any of the collected final positions, and repeat the process.
The exit rate $\aexit^{\traj}(t)$ which appears
 in Eq.~\eqref{eq:delta_aexit} is the rate at which the red
 sample trajectories leave the tube for the first time.
The small value $M=3$ is chosen here for illustration; to calculate exit
rates from experimental data, we use values of the order $10^4$, which are chosen dynamically,
see Appendix \ref{sec:algorithm} for details.
\textcolor{black}{Our algorithm assumes that the time series are Markovian;
we verify in Appendix \ref{app:parametrization} that this holds approximately for our experimental data.}
To demonstrate that 
 concatenating short measured trajectories
 does not artificially alter the dynamics,
we in 
Appendix \ref{sec:algorithm} 
vary the parameters of the algorithm, including $\RefillTime$ and $M$, 
and find that the inferred exit rate
is independent of the particular choices.
All trajectories shown here are actual experimental data, the reference path $\traj$
is the same as in Fig.~\ref{fig:extrapolated_rate} (a).
}
\end{figure}

\begin{figure*}[ht!]
\centering
	\includegraphics[width=\textwidth]{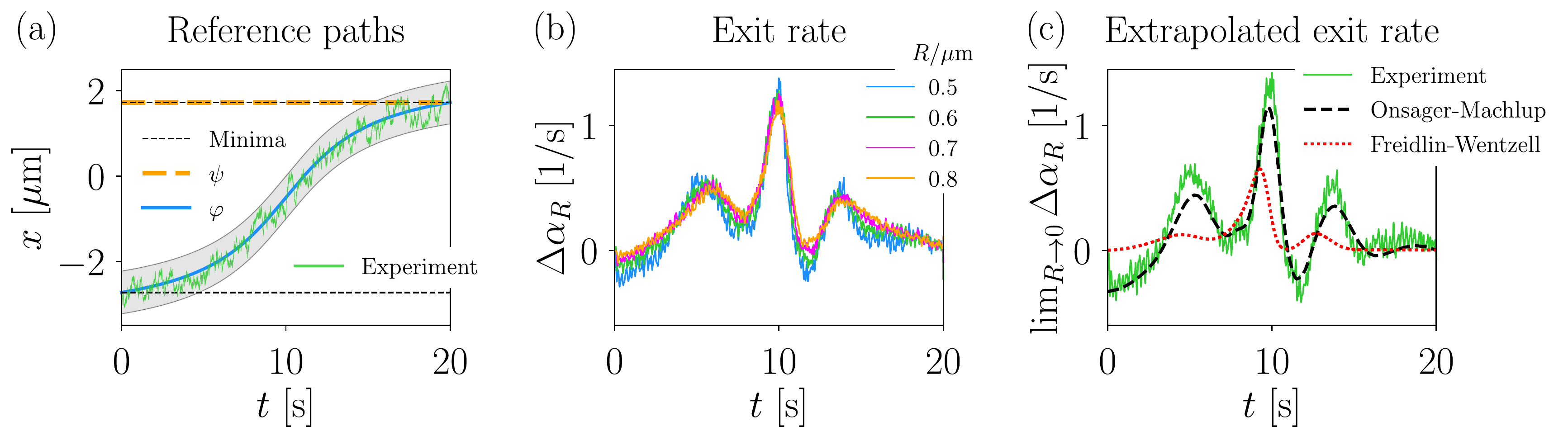}
\caption{
\label{fig:extrapolated_rate}
\textit{(a) Reference paths used to extract relative path probabilities.}
The thin horizontal dashed lines denote the minima in the experimental 
potential energy landscape, c.f.~Fig.~\ref{fig:setup_and_potential}; the 
upper dashed line additionally denotes the constant path $\trajTwo$,
as indicated by a thick orange dashed line.
The blue solid line denotes a path $\traj$ which moves from the left potential-energy minimum
 to the right minimum in 20 seconds.
The gray shaded region around the path $\traj$ indicates a tube of radius $R = 0.5\,\mu$m;
the green solid line depicts concatenated experimental time series, obtained using the
algorithm from Fig.~\ref{fig:illustration}.
\textit{(b)}
\textit{Exit rate differences for finite radius $R$.}
Colored solid lines denote the exit rate difference Eq.~\eqref{eq:delta_aexit},
 extracted directly from experimental time series for various values of the radius
 $R$, as indicated in the legend.
The used reference paths $\traj$, $\trajTwo$, are shown in subplot (a),
the shown exit rates are smoothed using
 a Hann window of width $0.1\,$s.
\textit{(c) Extrapolation of exit rate differences to radius $R=0$.} 
The green solid line denotes the extrapolation to $R = 0$ 
of (the pre-smoothing versions of) finite-radius exit rate differences as shown in subplot (b).
The shown extrapolated exit rate is smoothed using
 a Hann window of width $0.1\,$s.
The black dashed line and the red dotted line denote the difference in
 the OM and FW Lagrangians
for the paths $\traj$, $\trajTwo$, calculated using Eqs.~\eqref{eq:OnsagerMachlup},
\eqref{eq:FreidlinWentzell}, and the 
diffusivity and force estimated from the experimental data,
c.f.~Fig.~\ref{fig:setup_and_potential}.
}
\end{figure*}

 We now compare 
 the experimentally measured right-hand side of Eq.~\eqref{eq:LagrangianDifference}
 to the corresponding difference in  theoretical stochastic action Lagrangians.
From recorded
 experimental time series of a colloidal particle in a microchannel, 
as illustrated in Fig.~\ref{fig:setup_and_potential}, 
we evaluate 
 \begin{equation}
 \label{eq:delta_aexit}
 \Delta \aexit(t) \equiv \aexit^{\traj}(t) - \aexit^{\trajTwo}(t),
\end{equation}
for several finite values of $R$.
Since the sojourn probabilities for the paths and radii we consider are so
 small that not a single
recorded trajectory remains within the tube until the final time, we introduce a \textcolor{black}{cloning} algorithm
to obtain the experimental exit rate based on  concatenating short recorded trajectories, see
Fig.~\ref{fig:illustration} for an illustration and Appendix \ref{sec:algorithm}
for further details. 
For $\traj$ we consider a path which moves from the left minimum of the
experimental potential energy
to the right minimum of the potential energy
 in $\TimeDiff =20\,$s, as illustrated in Fig.~\ref{fig:extrapolated_rate} (a).
For $\trajTwo$ we consider a constant path, which rests at the right minimum for 
the duration $\TimeDiff =20\,$s,
 shown as the upper horizontal dashed line in Fig.~\ref{fig:extrapolated_rate} (a).
 \textcolor{black}{
This choice for $\trajTwo$ ensures
 that all time-dependence in the measured exit-rate differences
 can be attributed
to the exit rate pertaining to $\traj$.
In principle arbitrary pairs of twice-differentiable
 paths of the same duration can be considered in Eq.~\eqref{eq:ActionDifference};
 we illustrate this in Appendix~\ref{sec:algorithm}, where we show results for
 several other path pairs. 
 }
In Fig.~\ref{fig:extrapolated_rate} (b) we show the exit-rate differences Eq.~\eqref{eq:delta_aexit}
for the paths $\traj$, $\trajTwo$ for several finite values of $R$. 
We extrapolate
  to 
$R = 0$ as follows. 
Since the exit rate is invariant under a parity transformation around the instantaneous tube center $\traj(t)$,
for small radius the difference in exit rates scales as
	$\Delta \aexit(t) = \Delta \alpha^{(0)}(t) + R^2 \Delta \alpha^{(2)}(t) + \mathcal{O}(R^4)$.
 For every time $t$, we therefore fit a quadratic function
 	$f(t,R) = a(t) + R^2b(t)$
to measured finite-radius exit rates, as the ones shown in Fig.~\ref{fig:extrapolated_rate} (b), 
and extrapolate to vanishing radius as
	$\lim_{R \rightarrow 0} \Delta \aexit(t) \equiv a(t)$.
Figure \ref{fig:extrapolated_rate} (c) compares the result
 with the theoretical difference in OM Lagrangians
  \cite{onsager_fluctuations_1953,
stratonovich_probability_1971,durr_onsager-machlup_1978,fujita_onsager-machlup_1982},
\begin{equation}
\label{eq:OnsagerMachlup}
\mathcal{L}^{\traj}(t) = \frac{\gamma}{4k_{\mathrm{B}}T_0}\left[ \dot{\traj}(t) - \frac{1}{\gamma} F(\traj(t))\right]^2
+ \frac{1}{2 \gamma}  (\partial_x F)(\traj(t)),
\end{equation}
where 
\textcolor{black}{$\gamma = 1.75\cdot 10^{-8}\,\mathrm{kg}/\mathrm{s}$ 
is the friction coefficient, 
$k_{\mathrm{B}}T_0$ the thermal energy with $k_{\mathrm{B}}$
the Boltzmann constant and $T_0 = 294\,$K the absolute temperature at which the 
experiment is carried out,
and $F(x) = - (\partial_x U)(x)$ the force corresponding to the potential shown in Fig.~\ref{fig:setup_and_potential};
for  details on the parametrization of 
$\gamma$, $U(x)$,
see Appendix \ref{app:parametrization}.}
While in Fig.~\ref{fig:extrapolated_rate} (c) there are some minor
 differences between theory and experimental extrapolation, 
the overall agreement is very good.
This shows both that 
our protocol for extracting ratios of path probabilities from experiments
yields meaningful results, and confirms 
that relative path probabilities are indeed quantified by the OM Lagrangian \cite{stratonovich_probability_1971}. 
On the other hand, the difference in FW Lagrangians \cite{ventsel_small_1970}, 
  given by
\begin{equation}
\label{eq:FreidlinWentzell}
\mathcal{L}_{\mathrm{FW}}^{\traj}(t) = \frac{\gamma}{4k_{\mathrm{B}}T_0}\left[ \dot{\traj}(t) - \frac{1}{\gamma} F(\traj(t))\right]^2,
\end{equation}
and also shown in Fig.~\ref{fig:extrapolated_rate} (c),
disagrees considerably with both
the experimental data and the OM prediction.
This illustrates very clearly that in the context of physically observable
asymptotic sojourn probabilities at finite temperature, 
the two actions  Eqs.~\eqref{eq:OnsagerMachlup}, \eqref{eq:FreidlinWentzell} are not equivalent.
They differ by a term proportional to $\partial_x F$,
which in the nonlinear force profile we consider here 
 contributes considerably to the asymptotic exit rates.
\textcolor{black}{This highlights that, while 
in the context of the path-integral formalism,
the OM and FW actions are usually considered equivalent,
this is not the case for asymptotic sojourn probabilities.
We discuss this in more detail in Sect.~\ref{sec:path_integral_action} further below.}

\subsection{Most probable path from experiment}
The most probable path $\varphi^{*}$, also called instanton, 
connecting an initial point $\varphi^{*}(t_i) = x_i$ and a final point $\varphi^{*}(t_f) = x_f$, 
is given by
\begin{equation}
\label{eq:most_probable_path_def}
\varphi^{*} \equiv \lim_{R\rightarrow 0} \left[ \underset{\varphi}{\mathrm{argmin}} \int_{t_i}^{t_f} \mathrm{d}t\,\alpha_{R}^{\varphi}(t)\right],
\end{equation}
where we minimize over all continuous paths with given
 endpoints $\varphi(t_i) = x_i$, $\varphi(t_f) = x_f$.
This equation follows from maximizing the right-hand side of 
Eq.~\eqref{eq:ActionDifference} with respect to $\traj$
for any fixed $\trajTwo$,
and states that the most probable path
is the one where the exit rate diverges  slowest as $R \rightarrow 0$.

As in Fig.~\ref{fig:extrapolated_rate} we use $t_i = 0$, $t_f = 20\,$s, 
and for $x_i$, $x_f$,
consider the two minima of the experimental potential energy, c.f.~Fig.~\ref{fig:setup_and_potential}.
Using our experimental time series, 
we minimize the right-hand side of Eq.~\eqref{eq:most_probable_path_def},
but without the limit,
to obtain the most probable tube for  several finite values of $R$.
We subsequently extrapolate the result
to the limit $R \rightarrow 0$,
to obtain the most probable path;
for details see Appendix \ref{app:instanton_from_data}. 
In Fig.~\ref{fig:most_probable_path} (a) we compare the resulting experimental instanton
 to the directly minimized
  OM action, obtained by substituting Eq.~\eqref{eq:OnsagerMachlup} into Eq.~\eqref{eq:LagrangianDef}. 
 \begin{figure}[ht]
\centering
	\includegraphics[width=0.95\columnwidth]{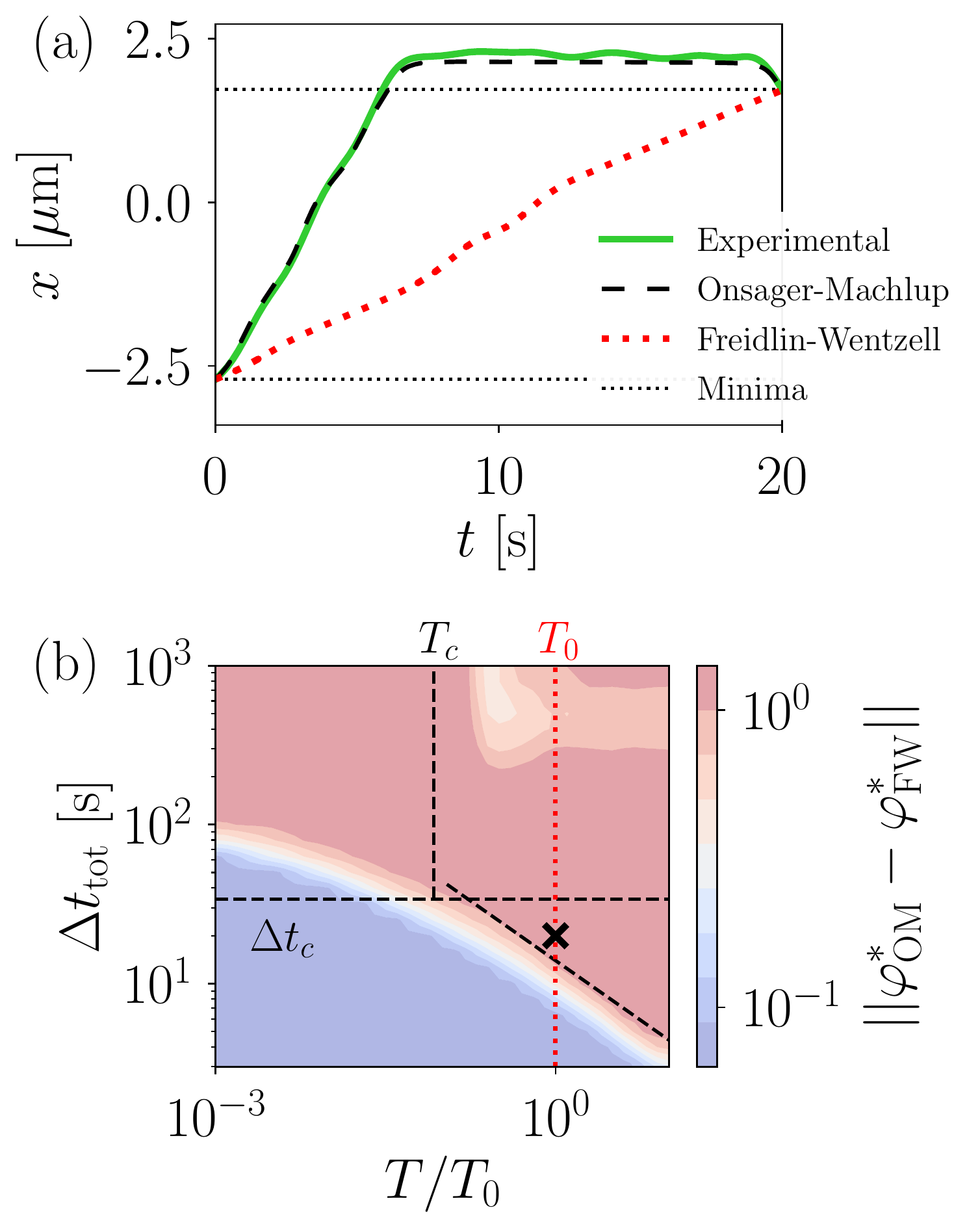}
\caption{
\label{fig:most_probable_path}
\textit{(a) Most probable paths for barrier crossing.} 
The green solid line denotes the most probable path
  extracted directly from experimental data, see Appendix \ref{app:instanton_from_data}
   for details.
The black dashed line is obtained by minimizing the integrated OM
 Lagrangian Eq.~\eqref{eq:OnsagerMachlup}, 
 the red dotted line is obtained by minimizing the integrated 
FW Lagrangian Eq.~\eqref{eq:FreidlinWentzell}.
\textit{(b) Mean difference between OM and FW instanton.}
 The actions corresponding to the Lagrangians 
 Eq.~\eqref{eq:OnsagerMachlup}, \eqref{eq:FreidlinWentzell}
 are minimized for various values of temperature $ T/ T_0$ and total transition time $\TimeDiff$. 
The plot shows the dimensionless  mean difference between the resulting instantons, 
as defined in Eq.~\eqref{eq:FW_OM_diff}; for technical details see Appendix \ref{app:calculation_of_phase_diagram}.
The horizontal and diagonal black dashed lines denote the 
crossover time $\Delta t_c \approx 34\,$s defined in Eq.~\eqref{eq:FWvsOM_1},
and the right-hand side of Eq.~\eqref{eq:FWvsOM_2}.
The black vertical dashed line denotes the crossover temperature
$ T/ T_0 \approx 0.08$ defined in Eq.~\eqref{eq:FWvsOM_3};
the red vertical dotted line denotes the reference temperature $T_0=294\,$K. 
The  cross denotes the parameters $( T/ T_0,\TimeDiff)$ used for subplot (a).
}
\end{figure}
As the figure shows, the extrapolated most probable tube agrees very 
well with the OM instanton,
demonstrating that the most probable path can be extracted 
from experimental data without fitting a model \textcolor{black}{for the stochastic dynamics}.
The FW instanton, obtained from minimizing the temporal integral over
the Lagrangian Eq.~\eqref{eq:FreidlinWentzell}, 
is also shown in Fig.~\ref{fig:most_probable_path} (a), 
and disagrees significantly with the experimental data.
\textcolor{black}{This confirms experimentally that the OM Lagrangian yields the correct action to describe 
physically observable most probable paths at finite temperature \cite{durr_onsager-machlup_1978,
adib_stochastic_2008}.}

\subsection{Range of validity of FW Lagrangian}
\textcolor{black}{
In the context of sojourn probabilities, the FW Lagrangian is derived for asymptotically
low noise \cite{ventsel_small_1970}, which in the present case means asymptotically low temperature.
Indeed, for fixed $\gamma$, $F$,
the first term in Eq.~\eqref{eq:OnsagerMachlup} scales with the inverse temperature, whereas
the second term is independent of temperature. 
One might thus expect that for sufficiently small temperature, the second term becomes irrelevant
and the Lagrangian Eq.~\eqref{eq:OnsagerMachlup} reduces to 
Eq.~\eqref{eq:FreidlinWentzell}.
To quantitatively understand in which parameter regime the FW Lagrangian predicts an instanton which is in agreement with the
 OM instanton, we now compare the minima of the functionals
obtained by substituting Eqs.~\eqref{eq:OnsagerMachlup}, \eqref{eq:FreidlinWentzell}
into Eq.~\eqref{eq:LagrangianDef},
for a wide range of 
 both the total duration  $\TimeDiff$ and 
 the temperature $T$.
For all temperatures, we 
use the friction coefficient $\gamma$ and the force profile $F$ 
inferred from our experimental data at the temperature $T_0 = 294\,K$.
In reality, $\gamma$ and $F$ of the experimental system shown in Fig.~\ref{fig:setup_and_potential} (a)
of course 
do depend on temperature \cite{peterman_laser-induced_2003}.}
\textcolor{black}{
We here hold those parameters fixed because
the focus of this section is not to understand the temperature-dependence of 
our particular experimental system, but of the theoretical instantons
predicted by Eqs.~\eqref{eq:OnsagerMachlup}, \eqref{eq:FreidlinWentzell}.
Note that, while in the FW Lagrangian Eq.~\eqref{eq:FreidlinWentzell}, the temperature 
only appears as an overall prefactor
which does not affect the variational extremum, 
varying the temperature in the OM Lagrangian Eq.~\eqref{eq:OnsagerMachlup}, 
 amounts to changing the relative size of the two terms.
}

In Fig.~\ref{fig:most_probable_path} (b) we show the numerically
 evaluated dimensionless average difference between FW and OM instanton,
\begin{equation}
\label{eq:FW_OM_diff}
	 || \traj_{FW}^{*} - \traj_{OM}^{*}||  \equiv \frac{1}{\TimeDiff \cdot L}  \int_{t_i}^{t_f} \mathrm{d}t ~\left| \traj_{FW}^{*}(t) - \traj_{OM}^{*}(t) \right|,
\end{equation}
as a function of the total duration $\TimeDiff = t_f - t_i$, and temperature $ T/T_0$, with the 
experimental temperature
$T_0 = 294\,K$ indicated in the plot by a red vertical dotted line.
For the typical length scale in Eq.~\eqref{eq:FW_OM_diff}  we use $L = 1\,\mu$m.
As can be seen, \textcolor{black}{
for short total duration $\TimeDiff$ there is a parameter regime where 
 the FW Lagrangian predicts the correct instanton.
 }
\textcolor{black}{
The FW Lagrangian has been derived as a low-temperature approximation
to the sojourn probability \cite{ventsel_small_1970}, 
but for total times $\TimeDiff \gtrsim 10^{2}\,$s 
the FW and OM instantons disagree even at the lowest temperature considered here;
we discuss the reason for this further below.
In Appendix~\ref{app:instantons_for_various_temperatures_and_durations} we show
 instantons from experimental and numerical data
for several more points in the $(T/T_0,\Delta t_{\mathrm{tot}})$ plane, all of which
 confirm that the OM instanton describes the measured instanton.
}

A quantitative estimate for 
the range of applicability of the FW action to determine the instanton is obtained by
 investigating for which parameters the second term in  Eq.~\eqref{eq:OnsagerMachlup}
is negligible as compared to the first term.
For this, we distinguish between the two limiting cases of short and long total duration.
If the total duration $\TimeDiff$ is short, the precise meaning of which will be quantified 
in the following,
 then the trajectory needs a large velocity to reach
the given final position $x_f = x_i + \Delta x$. 
For this scenario we estimate the typical velocity as $\dot{\traj} \approx \Delta x/ \TimeDiff$, 
so that $\dot{\traj}^2$ 
 dominates the first term in Eq.~\eqref{eq:OnsagerMachlup} if 
 the total time is much smaller than a crossover time
\begin{equation}
\label{eq:FWvsOM_1}
\TimeDiff \ll \Delta t_c \equiv \frac{ \gamma\,\Delta x}{\langle |F| \rangle},
 \end{equation}
 where the brackets $\langle ~\rangle$ denote a spatial average between
  $x_i$ and $x_f$, so that $\langle |F| \rangle$ represents the typical magnitude of the
  force between $x_i$ and $x_f$.
Using the experimentally inferred values for $\gamma$, $F$, Eq.~\eqref{eq:FWvsOM_1} 
yields $\Delta t_c \approx 34\,$s, which is 
shown in Fig.~\ref{fig:most_probable_path} (b) as horizontal dashed line.
For  a fast transition, $\TimeDiff \ll \Delta t_c$,
 the second term in the OM Lagrangian
 Eq.~\eqref{eq:OnsagerMachlup} is negligible
as compared to the first term
if 
\begin{align}
	\TimeDiff &\ll \frac{1}{\sqrt{ T/T_0}} \frac{\Delta x}{D_0 \sqrt{ \beta_0\langle | \partial_x F| \rangle}},
	\label{eq:FWvsOM_2}
\end{align}
where we write
 $\beta_0^{-1} \equiv k_{\mathrm{B}} T_0$ 
 as well as $D_0= k_{\mathrm{B}}T_0/\gamma$.
For our example system, 
$\Delta x/ (D_0 \sqrt{ \beta_0\langle | \partial_x F| \rangle}) \approx 14$\,s.
The right-hand side of Eq.~\eqref{eq:FWvsOM_2} is shown in Fig.~\ref{fig:most_probable_path} (b)
as diagonal black dashed line, 
and indeed in the lower left corner of the plot,
 where both Eqs.~\eqref{eq:FWvsOM_1}, \eqref{eq:FWvsOM_2},
are fulfilled, the FW action predicts the correct instanton.
Surprisingly, even though the FW action is in the context of sojourn probabilities
 only derived for low noise,
it predicts the correct instanton even at $ T/ T_0 = 1$ if the 
transition is fast enough; we illustrate this with an
 example in Appendix \ref{app:instantons_for_various_temperatures_and_durations}.
 \textcolor{black}{
For long total duration, $\TimeDiff \gg \Delta t_c$, we observe in 
Fig.~\ref{fig:most_probable_path} (b) that the FW instanton is different from the OM instanton even for the
smallest temperatures considered.
To understand this, we first discuss a simple scaling argument to determine 
at which
temperature the second term in Eq.~\eqref{eq:OnsagerMachlup} might be negligible,
 and then rationalize why even in this regime we observe deviations
between FW and OM instanton.}
For $\TimeDiff \gg \Delta t_c$, the first term in the
 OM Lagrangian Eq.~\eqref{eq:OnsagerMachlup} is expected to be
 of order $  \langle| F|\rangle^2/(4 \gamma k_{\mathrm{B}}T)$, so that the second term should be negligible if
  the temperature is much smaller than a crossover
 temperature $T_c$, defined by
\begin{align}
	\frac{ T}{ T_0} \ll \frac{ T_c}{ T_0} \equiv \frac{1}{2}  \frac{ \left\langle |\beta_0 F|^2\right\rangle }{ \left\langle | \beta_0\partial_x F| \right\rangle} .
	\label{eq:FWvsOM_3}
\end{align}
For our system, $T_c/T_0 \approx 0.08$, which is  shown as 
vertical dashed line in Fig.~\ref{fig:most_probable_path} (b);
\textcolor{black}{however, even in the
parameter regime where both $\TimeDiff \gg \Delta t_c$ and Eq.~\eqref{eq:FWvsOM_3}
are fulfilled, i.e.~in the upper left corner of Fig.~\ref{fig:most_probable_path} (b), 
the FW and OM instantons are distinct.
This is because
for long enough $\TimeDiff$ the OM instanton rests
close to one of the minima of the potential energy landscape,
as observed in
Fig.~\ref{fig:most_probable_path} (a).
That staying close to a potential minimum
leads to a particularly small exit rate (and hence large sojourn probability) 
is intuitively reasonable,
as the positive curvature of the potential
around the minimum  hinders particle exit from the tube.
Now if the trajectory rests close to a potential minimum, then both the velocity and the force are
very small, and hence the \textcolor{black}{second term in Eq.~\eqref{eq:OnsagerMachlup} 
may not be negligible as compared to the first term even for rather low temperatures}.
Thus, even at low temperature
 the FW action is not expected to produce the OM instanton whenever the latter 
 yields a path that temporarily rests close to a state where the deterministic force vanishes.}
\textcolor{black}{
In summary,  
 Fig.~\ref{fig:most_probable_path} (b) shows that  while 
 the FW instanton agrees with the OM instanton for short transitions, as 
 quantified by
 Eqs.~\eqref{eq:FWvsOM_1}, \eqref{eq:FWvsOM_2},
 a similar scaling argument for long transitions,
given by Eq.~\eqref{eq:FWvsOM_3},
fails.
This is because even at the lowest temperature
considered here, the second term in the OM action remains relevant for the instanton.
}

\section{Relation between the asymptotic sojourn probability
and the path-integral action}
\label{sec:path_integral_action}

\textcolor{black}{In the present section we show that the asymptotic-tube 
action, defined by Eq.~\eqref{eq:ActionDifference},
 is consistent with the path-integral formalism,
in the sense that both approaches lead to identical 
probability ratios for
 twice differentiable paths.} 

\textcolor{black}{
By formally writing the sojourn probabilities on
 the right-hand side of Eq.~\eqref{eq:ActionDifference} 
as path integrals, the equation reads
\begin{equation}
\label{eq:path_integral_limit}
 \frac{
 e^{-S[\varphi]}
 }{
 e^{-S[\psi]}
 }
 = 
\lim_{R \rightarrow 0}
\frac{ 
\int\mathcal{D}[X]\, \mathbb{1}_{R}^{\varphi}[{X}] 
e^{
	{-\SPI[X]}
	}
 }{
 \int\mathcal{D}[X] \,
 \mathbb{1}_{R}^{\psi}[{X}] 
 e^{
 	-\SPI[X]
	}
},
\end{equation}
where $S$ on the left-hand side denotes the OM action defined by Eqs.~\eqref{eq:LagrangianDef},
 \eqref{eq:OnsagerMachlup},
$\mathcal{D}[X]$ stands for a fictitious uniform measure on the space of all continuous paths \cite{williams_probability_1981}, 
$e^{{-\SPI[X]}}$ is the corresponding path weight which defines
the formal path-integral action $\SPI$,
$\mathbb{1}_{R}^{\varphi}[X]$
denotes the indicator function on the 
set of all paths that remain within a tube of radius $R$ around the reference path $\varphi$,
and for each path integral we consider
the initial point 
of the respective reference path as initial condition for the corresponding stochastic paths.
To give operational meaning to the symbolic path-integral expressions
 on the right-hand side of Eq.~\eqref{eq:path_integral_limit}, 
we interpret the expression as continuum limit of 
 the standard time-slicing procedure \cite{onsager_fluctuations_1953,
 haken_generalized_1976,
graham_path_1977,
dekker_functional_1978,
wissel_manifolds_1979,
langouche_functional_1979,
dekker_path_1980,
hunt_path_1981,
weber_master_2017,
wissel_manifolds_1979,
adib_stochastic_2008,
cugliandolo_rules_2017,
cugliandolo_building_2019}, 
as discussed in more detail below.
}

\textcolor{black}{
The limiting ratio of sojourn
 probabilities, Eq.~\eqref{eq:ActionDifference},
 defines 
the action $S$ on the left-hand side of Eq.~\eqref{eq:path_integral_limit};
 the path-integral action $\SPI$ in the integrand on the right-hand side of the equation
  is defined as a formal limit of 
a time-slicing procedure.
From Eq.~\eqref{eq:path_integral_limit}
one would assume that these two actions are equal: As $R$ on the right-hand side of 
the equation tends to zero, only the immediate neighborhood of the reference path remains
 within the tube. 
One would thus expect that
 the right-hand side tends to $e^{ - \SPI[\varphi]}/e^{ -\SPI[\psi]}$,
 which is equal to the left-hand side if $S= \SPI$, up to a possible
 path-independent additive contribution.}
\textcolor{black}{
However, 
the formal path-integral action on the right-hand 
side of Eq.~\eqref{eq:path_integral_limit} is not uniquely defined;
in fact, infinitely many equivalent stochastic actions can be employed,
including the OM and FW actions 
Eqs.~\eqref{eq:OnsagerMachlup}, \eqref{eq:FreidlinWentzell}
\cite{haken_generalized_1976,
wissel_manifolds_1979,
langouche_functional_1979,
hunt_path_1981,
adib_stochastic_2008}.
This is in sharp contrast to  
the limiting ratio of tube probabilities at finite temperature, 
which is described by the OM action 
Eq.~\eqref{eq:OnsagerMachlup},
as given on the left-hand side of Eq.~\eqref{eq:path_integral_limit} \cite{stratonovich_probability_1971,
durr_onsager-machlup_1978,ito_probabilistic_1978,
williams_probability_1981,fujita_onsager-machlup_1982,ikeda_stochastic_1989}.
}

\textcolor{black}{
To show how these two seemingly inconsistent 
mathematical results
are reconciled,
we now consider the usual time-slicing derivation of the 
path-integral action \cite{onsager_fluctuations_1953,
haken_generalized_1976,
dekker_functional_1978,
wissel_manifolds_1979,
hunt_path_1981,
adib_stochastic_2008,
weber_master_2017,
cugliandolo_building_2019}.
The key insight in the following discussion
 is that the continuum limit of the discretized path-integral action
 depends on whether the action is evaluated on a typical
realization of the Langevin equation, or on a differentiable
path.
The ambiguity in the stochastic action discussed in the literature \cite{haken_generalized_1976,
wissel_manifolds_1979,
langouche_functional_1979,
hunt_path_1981,
adib_stochastic_2008} only
emerges in the former case;
in the latter case, which is the appropriate scenario for comparing
the time-slicing action
to the limiting ratio Eq.~\eqref{eq:ActionDifference}, the OM action follows unambiguously,
i.e.~independent of the discretization scheme.}
\textcolor{black}{The relation between the action of rough and smooth paths
has been discussed from another perspective in previous works \cite{graham_path_1977,dekker_functional_1978}.}

\textcolor{black}{
For a one-dimensional stochastic process $X_t$, 
we consider the It\^{o}-Langevin equation
\begin{equation}
\label{eq:ItoLangevin}
\mathrm{d}X_t = a(X_t)\mathrm{d}t + b\,\mathrm{d}B_t,
\end{equation}
where $a(x) = D_0 \beta_0 F(x)$ is the drift, $\mathrm{d}B_t$ is the increment of the
 Wiener process, and the noise strength $b = \sqrt{ 2 D_0}$ is independent of position.
Since the noise is additive, the
 following derivation is independent of
whether one interprets Eq.~\eqref{eq:ItoLangevin} in the It\^{o} or
Stratonovich sense \cite{kloeden_numerical_1992,gardiner_stochastic_2009}.
}

\textcolor{black}{
We  consider a given continuous  path $Y_t$, $t \in [0,t_f]$,
 and aim to quantify its probability according to
the It\^{o}-Langevin Eq.~\eqref{eq:ItoLangevin}.
For this, we discretize the time interval $[0,t_f]$ into $N$ equally large slices
of duration $\Delta t = t_f/N$, and denote the position of the path 
at time $t_i \equiv i \cdot \Delta t$ by
$Y_i \equiv Y_{t_i}$.
Using the Markov property, we rewrite the joint probability density
 that a realization of the
Langevin Eq.~\eqref{eq:ItoLangevin},
after starting at $Y_0$ at time $t_0$,
is at the points $Y_i$ at times $t_i$, 
as
\begin{align}
\label{eq:time_discretized_path_integral}
&P(Y_N,t_N;Y_{N-1},t_{N-1};...;Y_1,t_1\mid Y_0,t_0) 
\\
&\quad= \prod_{i=0}^{N-1}P(Y_{i+1},t_{i+1}|Y_{i},t_{i}) 
\\
&\quad \equiv \mathcal{N}_N \exp\left[ - \sum_{i=0}^{N-1} \Delta \SPI_i\right],
\label{eq:time_discretized_path_integral2}
\end{align}
where
we define the normalization constant $\mathcal{N}_N \equiv \left( 2 \pi b^2 \Delta t\right)^{-N/2}$.
We now derive
 an expression for the discretized action
 $\sum_{i} \Delta \SPI_{i} \equiv \sum_i \Delta \SPI(Y_{i+1}, Y_i,\Delta t)$,
which is defined by 
Eq.~\eqref{eq:time_discretized_path_integral2},
 by calculating the short-time propagator of the Langevin dynamics.
Equation \eqref{eq:time_discretized_path_integral2} is the usual starting 
 point for the time-slicing derivation of the
 path integral \cite{onsager_fluctuations_1953,
haken_generalized_1976,
dekker_functional_1978,
wissel_manifolds_1979,
hunt_path_1981,
adib_stochastic_2008,
weber_master_2017,
cugliandolo_building_2019}.
}

\textcolor{black}{
At this point we do not require that the given continuous path
 $Y_t$ be a realization of the Langevin equation.
We only assume that  the increments
$\Delta Y_i \equiv Y_{i+1} - Y_i$ fulfill
$\Delta Y_i = \mathcal{O}(\Delta t^{1/2})$ for small $\Delta t$.
This condition holds
if $Y_t$ is a typical realization of the Langevin equation,
but it is also true
if $Y_t$ is a continuously differentiable path, for which in fact 
the stronger condition $\Delta Y_i = \mathcal{O}(\Delta t)$ holds.
}

\textcolor{black}{
For the It\^{o}-Langevin Eq.~\eqref{eq:ItoLangevin},
the increment 
$\Delta X \equiv X_{\Delta t} - X_{0}$
for a short time interval $\Delta t$, and with initial condition $X_0$, 
follows via an It\^{o}-Taylor expansion as \cite{kloeden_numerical_1992},
 \begin{align}
 \nonumber
\Delta X  &=  \Delta t^{1/2}\,b\,\Delta W + \Delta t\,a(X_0) +  \Delta t^{3/2}\,a'(X_0) b\Delta Z 
\\ &\quad
+ \mathcal{O}(\Delta t^{2}),
 \label{eq:order_15_discretization}
 \end{align}
 where  
 $a'$ 
  denotes the spatial derivative of the drift $a$, and
 where
  $(\Delta W,\Delta Z)$ are distributed according to a two-dimensional Gaussian distribution \cite{kloeden_numerical_1992}
  \begin{equation}
  P(\Delta W, \Delta Z) = \frac{1}{2 \pi \sqrt{ \det{\Sigma}  }} e^{ - \frac{1}{2} (\Delta W, \Delta Z) \cdot \Sigma^{-1} \cdot (\Delta W, \Delta Z)^T},
  \end{equation}
  with 
  \begin{align}
  \Sigma = \begin{pmatrix} 1 & 1/2 \\  1/2 & 1/3
  \end{pmatrix}.
  \end{align}
By $\mathcal{O}(\Delta t^{2})$, we in Eq.~\eqref{eq:order_15_discretization}
subsume both random and and deterministic
 terms that scale at least as $\Delta t^{2}$ for small $\Delta t$.
}

\textcolor{black}{
The probability density 
 for
 observing
  the $i$-th increment $\Delta Y_i$ of the given continuous path $Y_t$
in a realization of the Langevin dynamics
 is now obtained as \cite{kampen_stochastic_2007}
\begin{align}
\label{eq:prob_def}
& P(Y_{i+1},t_{i+1}\mid Y_i,t_{i}) \equiv  
 \int\mathrm{d}\Delta W\int\mathrm{d}\Delta Z \,
\\ & \quad \times
 \left[\vphantom{\frac{1}{2}}\delta\left(\Delta Y_i - \Delta X(\Delta W, \Delta Z)\right) P(\Delta W, \Delta Z)\right]\,,
\nonumber
\end{align}
where the increment as a function of the noise, $\Delta X(\Delta W, \Delta Z)$,
 is given by Eq.~\eqref{eq:order_15_discretization}, with initial condition $X_0 \equiv Y_i$,
and $\delta$ is the Dirac-delta distribution.
The delta distribution can be used to directly evaluate the integral over $\Delta W$;
 the remaining integral over $\Delta Z$ in Eq.~\eqref{eq:prob_def} is a Gaussian integral, 
which evaluates to
\begin{align}
P(Y_{i+1},t_{i+1}\mid Y_i,t_{i})
 &= \frac{1}{\sqrt{ 2\pi b^2 \Delta t}} e^{ - \Delta \SPI_i},
\end{align}
where
\begin{align}
\label{eq:DS_discrete}
\Delta \SPI_i &\equiv \Delta \SPI(Y_{i+1}, Y_i,\Delta t) 
\\ &\equiv \frac{\Delta t }{2b^2} \left( \frac{\Delta Y_i}{\Delta t} 
- a(Y_i) \right)^2 
- \frac{a'(Y_i)}{2b^2} \left( \Delta Y_i^2 - b^2\Delta t\right)
\nonumber
\\ & \qquad  + \mathcal{O}(\Delta t^{3/2}).
\label{eq:DS_discrete_}
\end{align}}\textcolor{black}{
From the appearance of $a'(Y_i)$ in Eq.~\eqref{eq:DS_discrete_} it is apparent
 why we consider Eq.~\eqref{eq:order_15_discretization}
beyond linear order in the time increment: the term of order $\Delta t^{3/2}$ in the
  discretized Langevin equation
  in fact
contributes a term of order $\Delta t$ in the short time propagator.}

\textcolor{black}{
In Eq.~\eqref{eq:DS_discrete_},
the drift $a$ and its derivative are evaluated at the initial point of each time slice.
Equivalently, we can use any other point in the interval $[Y_i, Y_{i+1}]$, and we parametrize
the choice by a parameter $\dparam \in[0,1]$ as
$\bar{Y}^{\dparam}_i\equiv  Y_i  + \dparam \left ( Y_{i+1} - Y_i\right)$.
Taylor expanding $a$, $a'$ around $\bar{Y}^{\dparam}_i$, 
substituting the result into Eq.~\eqref{eq:DS_discrete}, and rearranging, we obtain
\begin{align}
\label{eq:DS_discrete2}
\Delta \SPI_i &\equiv \frac{\Delta t }{2b^2} \left( \frac{\Delta Y_i}{\Delta t} 
- a(\bar{Y}^{\dparam}_i) \right)^2 
\\ & \qquad 
- \frac{a'(\bar{Y}^{\dparam}_i)}{2b^2} \left[ (1-2 \dparam) \Delta Y_i^2 - b^2\Delta t\right] + \mathcal{O}(\Delta t^{3/2}).
\nonumber
\end{align}
where we use that $\Delta Y_i = \mathcal{O}(\Delta t^{1/2})$.
This equation is, to order $\Delta t$, equivalent to Eq.~\eqref{eq:DS_discrete},
which is recovered from Eq.~\eqref{eq:DS_discrete2} 
by Taylor expanding $a$, $a'$ around $Y_i$ again.
Therefore,  despite the explicit appearance of $\dparam$ in Eq.~\eqref{eq:DS_discrete2},
the expression is to order $\Delta t$
 independent of this parameter \cite{haken_generalized_1976,
wissel_manifolds_1979,
langouche_functional_1979,
adib_stochastic_2008}.
}

\textcolor{black}{
Equations \eqref{eq:time_discretized_path_integral2}, \eqref{eq:DS_discrete2},
describe  a $N$-dimensional probability density,
which can be evaluated on any continuous path that 
obeys $\Delta Y_i = \mathcal{O}(\Delta t^{1/2})$.
Since the sum over the $\Delta \SPI_i$ in Eq.~\eqref{eq:time_discretized_path_integral2} 
has $N$ terms,
the sum over the
terms of order $\Delta t^{3/2} \sim N^{-3/2}$
from Eq.~\eqref{eq:DS_discrete2},
is of order $\mathcal{O}(N^{-1/2})$.
Thus, while the parameter $\dparam$ appears
explicitly in Eq.~\eqref{eq:DS_discrete2}, 
the discretized action in the probability density Eq.~\eqref{eq:time_discretized_path_integral}
depends on $\dparam$ only to subleading order $\mathcal{O}( N^{-1/2})$, which
becomes irrelevant in the limit $N \rightarrow \infty$.
Equations \eqref{eq:time_discretized_path_integral2}, \eqref{eq:DS_discrete2},
thus
constitute a 1-parameter family, parametrized by $\dparam \in [0,1]$, 
of asymptotically equivalent expressions
for the $N$-point probability density,
evaluated on a given continuous path
that obeys $\Delta Y_i = \mathcal{O}(\Delta t^{1/2})$ \cite{haken_generalized_1976,
wissel_manifolds_1979,
langouche_functional_1979,
adib_stochastic_2008}.
}

\textcolor{black}{
We first consider the formal continuum limit $N \rightarrow \infty$ of the exponent
in Eq.~\eqref{eq:time_discretized_path_integral2} for
 the case where $Y_t$ is a realization of the Langevin 
Eq.~\eqref{eq:ItoLangevin},
which we denote by $Y_t \equiv X_t$.
For a realization of the Langevin equation, 
 the formal continuum limit of the action defined by Eqs.~\eqref{eq:time_discretized_path_integral2}, 
\eqref{eq:DS_discrete2} follows 
as
\begin{align}
\label{eq:continuum_limit_brownian}
\SPI[X] &\equiv \lim_{N \rightarrow \infty} \sum_{i=0}^{N-1} \Delta \SPI(X_{i+1},X_i,\Delta t) 
\\ &=  \int_0^{t_f}\mathrm{d}t
\left[ \frac{1}{2b^2}
 \left( \dot{X}_t - a(X_t) \right)^2 + \dparam \,a'(X_t)\right]\,,
 \label{eq:continuum_limit_brownian2}
\end{align}
see Appendix~\ref{app:convergence} for a derivation of the second term in
 Eq.~\eqref{eq:continuum_limit_brownian2} from Eq.~\eqref{eq:DS_discrete2}.
For special choices of $\dparam$, Eq.~\eqref{eq:continuum_limit_brownian2} yields
 the Freidlin-Wentzell ($\dparam=0$),
Onsager-Machlup ($\dparam = 1/2$), or H\"anggi-Klimontovich ($\dparam = 1$) action.
That the formal continuum limit Eq.~\eqref{eq:continuum_limit_brownian2} depends on
the parameter $\dparam$ is precisely the well-studied 
fact
that the formal path-integral action is not uniquely defined \cite{haken_generalized_1976,
wissel_manifolds_1979,
langouche_functional_1979,
adib_stochastic_2008}.
In practice, this is not an issue:
As it stands,
the formal expression Eq.~\eqref{eq:continuum_limit_brownian2} 
is not mathematically well-defined.
More explicitly, for
 a  realization of the Langevin equation it holds
that
 \begin{align}
 \label{eq:diverging_term}
\frac{1}{2b^2}\int\mathrm{d}t\,\dot{X}_t^2 &\equiv \frac{1}{2b^2} \lim_{N \rightarrow \infty}\sum_{i=0}^{N-1} \Delta t \,\frac{\Delta X_i^2}{\Delta t^2} 
\\ & 
=
\frac{1}{2}\lim_{N \rightarrow \infty} \left[\sum_{i=0}^{N-1} \Delta W_i^2 + \mathcal{O}({N}^{-1/2})\right], 
 \label{eq:diverging_term2}
 \end{align}
 where we use
$ \Delta X_i^2/\Delta t = b^2\Delta W_i^2  + \mathcal{O}(\Delta t^{3/2})
 =b^2\Delta W_i^2 + \mathcal{O}(N^{-3/2})$,
which follows from  Eq.~\eqref{eq:order_15_discretization}.
Because  $\langle \Delta W_i^2 \rangle = 1$ and all the noise increments $\Delta W_i$ are
independent, 
the expected value of the limit Eq.~\eqref{eq:diverging_term2} is infinite.
Consequently, Eq.~\eqref{eq:continuum_limit_brownian2} 
should be interpreted in its discretized
 form Eq.~\eqref{eq:DS_discrete2}, 
 which is to leading order independent of $\dparam$ \cite{haken_generalized_1976,
wissel_manifolds_1979,
langouche_functional_1979,
adib_stochastic_2008}.
Furthermore, as we detail in Appendix~\ref{app:convergence},
if the diverging square term Eq.~\eqref{eq:diverging_term} is subtracted 
from the discretized action before taking the limit $N\rightarrow \infty$, 
 a well-defined expression is obtained, given by the Girsanov formula \cite{cameron_transformations_1944,girsanov_transforming_1960,oksendal_stochastic_2007}
and 
 independent of the value of $\dparam$ used in 
Eq.~\eqref{eq:DS_discrete2}.
}

\textcolor{black}{
Secondly, we consider the continuum limit $N \rightarrow \infty$ of the exponent
in Eq.~\eqref{eq:time_discretized_path_integral2} for
 the case where $Y_t$ is a continuously
  differentiable path with square-integrable derivative,
which we denote by $Y_t \equiv \varphi_t$.
While any such path occurs with probability zero as realization of
 the Langevin equation,
it is of course possible to
evaluate the $N$-point probability density Eq.~\eqref{eq:time_discretized_path_integral}
on a given set of positions  parametrized by  $\varphi_t$.
For a continuously differentiable path we have $\Delta \varphi_i = \mathcal{O}(\Delta t)$,
so that 
 $\Delta Y_i^2 \equiv \Delta \varphi_i^2 = \mathcal{O}(\Delta t^2)$,
 and hence 
  the continuum limit
 of Eqs.~\eqref{eq:time_discretized_path_integral2}, 
\eqref{eq:DS_discrete2}, 
 follows as
\begin{align}
\label{eq:continuum_limit_smooth}
\SPI[\varphi] &\equiv \lim_{N \rightarrow \infty} \sum_{i=0}^{N-1} \Delta \SPI( \varphi_{i+1},\varphi_i,\Delta t) 
\\ &=  \int_0^{t_f}\mathrm{d}t
\left[ \frac{1}{2b^2}
 \left( \dot{\varphi}_t - a(\varphi_t) \right)^2 + \frac{1}{2}\,a'(\varphi_t)\right]\,.
 \label{eq:continuum_limit_smooth2}
\end{align}
In contrast to Eq.~\eqref{eq:continuum_limit_brownian}, this limit is independent of 
the choice of $\dparam$, and always given by the OM action
Eqs.~\eqref{eq:LagrangianDef}, \eqref{eq:OnsagerMachlup} \textcolor{black}{\cite{graham_path_1977}},
which is seen by substituting $a(x) = F(x)/\gamma$, $b = \sqrt{ 2 \kT_0/\gamma}$
 into
Eq.~\eqref{eq:continuum_limit_smooth2}.
Furthermore,
because the path ${\varphi}_t$ is continuously
 differentiable with square-integrable derivative, 
the integral over $\dot{\varphi_t}^2$ is finite,
and hence
the functional Eq.~\eqref{eq:continuum_limit_smooth2}
 is well-defined. }
 
 \textcolor{black}{
 The two continuum limits Eqs.~\eqref{eq:continuum_limit_brownian2}, 
 \eqref{eq:continuum_limit_smooth2}
show  how it is mathematically consistent that the formal path-integral action is not unique \cite{haken_generalized_1976,
wissel_manifolds_1979,
langouche_functional_1979,
adib_stochastic_2008},
  whereas the limiting ratio Eq.~\eqref{eq:ActionDifference} is \cite{stratonovich_probability_1971,
durr_onsager-machlup_1978,ito_probabilistic_1978,
williams_probability_1981,fujita_onsager-machlup_1982,ikeda_stochastic_1989}.
  More explicitly, the ambiguity in the continuum limit of the discretized
   action Eq.~\eqref{eq:time_discretized_path_integral2}, \eqref{eq:DS_discrete2}
is only observed if the discretized action is evaluated
 on a typical realization of the Langevin Eq.~\eqref{eq:ItoLangevin},
in which case the formal expression Eq.~\eqref{eq:continuum_limit_brownian2} is obtained.
However, 
in Eq.~\eqref{eq:ActionDifference} we consider two twice differentiable reference paths,
for which
the limiting ratio of $N$-point probability densities follows 
from Eq.~\eqref{eq:continuum_limit_smooth2}
unambiguously as
\begin{equation}
\label{eq:OM_via_path_integrals}
\lim_{N\rightarrow \infty}\frac{P(\varphi_{N},t_{N};...;\varphi_1,t_1\mid \varphi_0,t_0)}{P(\psi_{N},t_{N};...;\psi_1,t_1\mid \psi_0,t_0)} = \frac{e^{-S[\varphi]}}{e^{-S[\psi]}},
\end{equation}
which is identical to the limiting ratio of tube probabilities Eq.~\eqref{eq:ActionDifference}
\cite{stratonovich_probability_1971,
durr_onsager-machlup_1978,ito_probabilistic_1978,
williams_probability_1981,fujita_onsager-machlup_1982,ikeda_stochastic_1989}.
}

\section{Discussion}

\subsection{Summary of results}

In this work, we establish a protocol to determine ratios of path probabilities
 from measured time series, without fitting a stochastic model to the data.
Applying this protocol to time series of a colloidal particle in a microchannel,
we find that the Onsager-Machlup action Lagrangian \cite{onsager_fluctuations_1953,
stratonovich_probability_1971,durr_onsager-machlup_1978,fujita_onsager-machlup_1982}
describes both ratios of
path probabilities and the most probable path extracted from our experimental data\textcolor{black}{,
 validating classical theoretical results \cite{stratonovich_probability_1971,
durr_onsager-machlup_1978,ito_probabilistic_1978,
williams_probability_1981,fujita_onsager-machlup_1982,ikeda_stochastic_1989}.}
The Freidlin-Wentzell action \cite{ventsel_small_1970,touchette_large_2009,grafke_numerical_2019} 
disagrees with our \textcolor{black}{finite-temperature} experimental results, and we quantify
for which parameters it is expected to predict the \textcolor{black}{correct most probable path}.
\textcolor{black}{By careful analysis of the continuum limit of the usual time-slicing approach to the path-integral formalism,
we resolve the apparent inconsistency that the formal path-integral action is not unique, 
whereas
both theoretical and experimental results point to a definitive 
probability ratio for differentiable paths.}

Our results 
 constitute a \textcolor{black}{model-free} experimental measurement of relative likelihoods
of stochastic trajectories, 
and demonstrate that from a physical point of view 
there is no ambiguity 
as to which stochastic action
describes observable relative path probabilities\textcolor{black}{,
as defined in Eq.~\eqref{eq:ActionDifference}.}

\subsection{\textcolor{black}{Probability vs.~probability density
}
}

\textcolor{black}{The key idea in our approach is to measure finite-radius sojourn probabilities, 
and to extrapolate the results to
the vanishing-radius limit using Eq.~\eqref{eq:ActionDifference}.
Our measurements hence focus on \textit{probabilities},
as opposed to \textit{probability densities},
which is because only non-zero probabilities,
such as the sojourn probability at finite radius,
can be directly measured in an experiment.
}

\textcolor{black}{
For overdamped Langevin dynamics, 
the sojourn probability for small-but-finite radius can be calculated as \cite{stratonovich_probability_1971,
 ito_probabilistic_1978,
 kappler_stochastic_2020}
\begin{align}
\label{eq:asymptotic_sojourn}
P_R^{\varphi}(t_f) = P_R^{\mathbb{W}}(t_f) \left[ e^{-S[\varphi]} + \mathcal{O}(R)\right],
\end{align}
where $S[\varphi]$ is the OM action defined by Eqs.~\eqref{eq:LagrangianDef}, 
\eqref{eq:OnsagerMachlup},
and where $P_R^{\mathbb{W}}(t_f)$ is the probability that a rescaled Wiener process,
starting at the origin and with the same increment-variance as the random force in the Langevin dynamics,
remains within a tube of radius $R$ until the final time $t_f$.
Substituting the asymptotic sojourn probability Eq.~\eqref{eq:asymptotic_sojourn}
for both $\varphi$, $\psi$ in Eq.~\eqref{eq:ActionDifference},
and noting that $P_R^{\mathbb{W}}(t_f)$ is for additive noise
independent of the reference path \cite{stratonovich_probability_1971,
 ito_probabilistic_1978,
 kappler_stochastic_2020},
the left-hand side of Eq.~\eqref{eq:ActionDifference} is obtained.
In that sense our experiments directly probe the measure
induced on the space of all continuous paths 
by the observed stochastic dynamics.}
\textcolor{black}{
For overdamped Langevin dynamics, Eq.~\eqref{eq:ActionDifference} is straightforwardly generalized to higher dimensions,
where a tube consists of a ball of radius $R$ around the
 reference path \cite{stratonovich_probability_1971,
 ito_probabilistic_1978,
 kappler_stochastic_2020}.
}

\textcolor{black}{
Following Ref.~\cite{durr_onsager-machlup_1978},
Eq.~\eqref{eq:asymptotic_sojourn} can be compared to the
probability that a one-dimensional real random variable $X$ is within a small interval
$I^x_{R} \equiv [x-R,x+R]$, $P(X \in I^x_{R})$. 
If the probability measure associated with $X$ 
has a density $\rho(x)$ with respect to the usual Lebesgue measure $\lambda$, then 
the probability can be written as
\begin{align}
\label{eq:one_dim_density}
P(X \in I_{R}^{x}) &=
  \lambda(I_{R}^{x}) \left[ \rho(x) + \mathcal{O}(R)\right],
\end{align}
where, due to the translation invariance of the Lebesgue measure, 
$\lambda(I_{R}^{x}) = |[x-R,x+R]| = 2R$ is independent of $x$.}

\textcolor{black}{
Comparing Eqs.~\eqref{eq:asymptotic_sojourn}, 
\eqref{eq:one_dim_density}, one might be inclined 
to think of $e^{-S[\varphi]}$ as an (unnormalized) probability density 
 with respect to the Wiener measure.
It is, however, apparent that this cannot be true \cite{durr_onsager-machlup_1978}, because
 the OM functional is not even well-defined for a typical 
realization of the Langevin equation, c.f.~Eq.~\eqref{eq:diverging_term2}.
 While the measure induced 
 on the space  of all continuous paths 
 by overdamped additive-noise Langevin dynamics
 does has a density with respect to the Wiener measure,
this density
 is not described by the OM functional, but rather
  by the Girsanov formula \cite{oksendal_stochastic_2007,
  cameron_transformations_1944,
  girsanov_transforming_1960}, c.f.~Appendix~\ref{app:convergence}.
}

\subsection{Extension to other types of stochastic dynamics}

\textcolor{black}{
In general, 
 for any reaction coordinate $X_t$, 
 independent of the precise nature of its stochastic
dynamics, the probability to remain within an asymptotically small tube 
is the experimentally relevant characterization of its path probabilities.
After all, it is precisely realizations of $X_t$ that one observes,
and the
asymptotic sojourn probability very concretely characterizes how 
these realizations behave.
This means that the sojourn probability is also a physically relevant observable
 for other variants of Langevin dynamics, which are  typically
obtained from projecting high-dimensional dynamics onto 
a low-dimensional reaction coordinate \cite{zwanzig_memory_1961,mori_transport_1965}.
If there is time-scale separation between fast orthogonal degrees of freedom
and the reaction coordinate, then such a projection leads to a Langevin equation with
multiplicative noise \cite{berezhkovskii_time_2011}, i.e.~configuration-dependent diffusivities.
If there is no time-scale separation, 
then the reaction coordinate is described by non-Markovian
Langevin dynamics.}
\textcolor{black}{Physical examples  
where memory effects are relevant include 
 conformational transitions in small molecules \cite{zuckerman_transition_2002,
 min_observation_2005,daldrop_butane_2018},
 colloidal particles in solution on very short time scales \cite{franosch_resonances_2011,
 daldrop_external_2017,
 kowalik_memory-kernel_2019},
 or the motion of cells \cite{selmeczi_cell_2005,
 mitterwallner_non-markovian_2020}.
}

\textcolor{black}{
Whether a limiting process as described in Eq.~\eqref{eq:ActionDifference}
leads to a finite result 
for an arbitrary reaction coordinate $X_t$ 
depends on the details of its stochastic dynamics.
For example, for overdamped Langevin dynamics with multiplicative noise,
D\"urr and Bach have shown that the limit in general does not exist \cite{durr_onsager-machlup_1978}.
This is conceptually similar to the fact that, also on  finite-dimensional spaces,
 not every physically relevant
 probability measure has a probability \textit{density} with
respect to the Lebesgue measure.
However, 
 even if the limiting ratio of sojourn probabilities is not finite,
 which means that one asymptotically small tube is infinitely more
  likely than the other, 
 the  sojourn probability for \textit{small-but-finite} radius is still an experimentally 
accessible quantity that describes the behavior of a reaction coordinate \cite{kappler_sojourn_2020}.
}

\textcolor{black}{Theoretically calculating the finite-radius sojourn probability
for a given stochastic dynamics, reference path, and small-but-finite tube radius
is a conceptually straightforward mathematical task:
This is an absorbing-boundary problem with moving boundaries for the reaction coordinate $X_t$ 
\cite{kappler_stochastic_2020,kappler_sojourn_2020}.
Whether absorbing boundaries should also be introduced for 
 orthogonal degrees of freedom,
such as the velocity in
 the case of inertial Langevin dynamics, or more generally memory degrees of freedom,
will in general depend on the physical  question under investigation.
 }

\textcolor{black}{
From an experimental point of view, 
 measuring the sojourn probability 
 of an observable
at a small finite radius tube radius
is also a well-defined problem.
As we do in this work, one needs to count
 which fraction of the recorded trajectories
 (all of which start inside the tube) leave
 the tubular neighborhood around the reference 
path for the first time at each recorded time.
One key aspect of our algorithm is how we overcome the
exponential temporal decay of the number of sample trajectories that have never left the tube:
We periodically increase the number of sample trajectories at each multiple of $\Delta \mathcal{T}$,
by
drawing new measured trajectories with initial conditions
according to the
instantaneous distribution 
of the current trajectories inside the tube.
As it only assumes Markovianity, this algorithm is directly 
applicable 
also for overdamped 
Langevin dynamics with multiplicative noise.
If the dynamics of the reaction coordinate is non-Markovian
our algorithm needs to be amended. 
A possible extension would be to not only use the instantaneous distribution of trajectories
 at multiples of $\Delta \mathcal{T}$ to draw new sample trajectories.
Instead, at iteration $k$ a multiple-time distribution could be considered, which
 includes the positions of
 trajectories at several times between
 $k \cdot \Delta \mathcal{T} - \tau$ and
 $k \cdot \Delta \mathcal{T}$,
where  $\tau$ denotes the longest memory time scale of the system.
A newly drawn trajectory would then need to approximately share
the same recent history  with a current trajectory,
so that the two might be considered at approximately
the same non-Markovian state.}

\textcolor{black}{The dynamics of the system we consider in the present work
is time-homogeneous,
meaning that both the external force, and the strength of the random force, 
are independent of time.}
\textcolor{black}{
While theoretically calculating the finite-radius sojourn probability 
 for a time-dependent force protocol
 is a conceptually clear task \cite{kappler_stochastic_2020,
kappler_sojourn_2020},
extending our data analysis to time-dependent forces means that more data needs to be collected.
More explicitly, in concatenating measured trajectories, as illustrated in Fig.~\ref{fig:illustration},
one needs to make sure that the external and random forces consistently follow 
the time-dependent protocol.
In practice this can be achieved by measuring a large number of independent 
experimental trajectories, 
each with the given protocol,
and only considering the value of each trajectory at the current state of the protocol 
in the cloning algorithm.
This should be contrasted with the algorithm in Appendix~\ref{sec:algorithm}, 
where
the time $t$ that appears in the sojourn probability 
does not need to agree with the physical time that has passed
since the recording of a trajectory started.
}

\textcolor{black}{
Both the extrapolated exit rate and the theoretical model used for
predictions
 in Fig.~\ref{fig:extrapolated_rate} (c) 
are based on measurements pertaining to comparable
length- and time scales,
and on which
the motion
of the colloidal particle
is  well-described by an overdamped Langevin model.
This is why even in the limit of asymptotically small tube radius, 
inertial effects do not play a role for our results.
If one were able to experimentally observe the dynamics 
in the colloidal-particle system shown in Fig.~\ref{fig:setup_and_potential} (b)
with arbitrary precision,
one would inevitably observe 
inertial effects or noise correlations on some small scale \cite{franosch_resonances_2011,lukic_motion_2007}.
However, any measurement can of course only directly probe the physics down to scales 
that are resolvable by the measurement apparatus.
This means that, very generally, a  limiting ratio for the sojourn probability
obtained directly from data should always be compared
to a theory that describes the dynamics on the scale 
 used for measuring the finite-radius exit rate.
That any measurement fixes a modeling scale
 is of course not a deficiency of
our particular approach, but a fundamental property of the physical sciences.
}

\subsection{
\textcolor{black}{Entropy production and further applications}
}

\textcolor{black}{
Equations \eqref{eq:LagrangianDifference} and
\eqref{eq:most_probable_path_def} relate path properties to experiment
by formulating them as vanishing-radius limits of 
tube properties.
This is a general strategy to relate questions
on individual paths to measurement:
If a given single-path statement 
can be reformulated in terms of a limit of tubes, 
the resulting finite-radius expression is accessible in measurement.
The corresponding finite-radius observables
can then be either investigated directly, 
or used to infer a vanishing-radius limit.} \textcolor{black}{
This method to experimentally access single-path properties
 will be particularly important for the field of stochastic thermodynamics, 
which extensively employs the concept of individual trajectories \cite{sekimoto_stochastic_2010,
seifert_stochastic_2012}.
For example, entropy production is in this context typically quantified by the probability ratio 
of forward and backward paths \cite{seifert_stochastic_2012}, 
which can be accessed experimentally by
considering 
ratios of sojourn probabilities for forward- and backward finite-radius tubes \cite{kappler_irreversibility_2020}.
}

\textcolor{black}{
In this study we probe the probability distribution on the space of all trajectories at finite temperature. However, our measurement algorithm is directly applicable also in the low-noise limit\textcolor{black}{, as considered in Freidlin-Wentzell theory \cite{ventsel_small_1970,touchette_large_2009}}, or in the study of switching transitions, 
where the full probability distribution in path space, induced by Langevin dynamics,
may be very concentrated around a single path
\cite{dykman_optimal_1992,
nickelsen_anomalous_2018,
luchinsky_analogue_1998,
chan_paths_2008,
herbert_predictability_2017}.}
\textcolor{black}{
One concrete example for possible future research on switching transitions are
 conformational changes of small molecules \cite{zuckerman_transition_2002,daldrop_butane_2018}.
For such molecules, measuring sojourn probabilities of dihedral angle reaction coordinates
 will allow to quantitatively study transitions between metastable states, without the need
to fit a stochastic model to the data.}

\textcolor{black}{
In Sect.~\ref{sec:path_integral_action} we have shown that
 the time-slicing approach to path integrals \cite{haken_generalized_1976,
wissel_manifolds_1979,
langouche_functional_1979,
adib_stochastic_2008},
is 
consistent with the mathematical results on the asymptotic ratio of sojourn probabilities, Eq.~\eqref{eq:ActionDifference} \cite{stratonovich_probability_1971,
durr_onsager-machlup_1978,ito_probabilistic_1978,
williams_probability_1981,fujita_onsager-machlup_1982,ikeda_stochastic_1989,
kappler_stochastic_2020},
in the sense that both unambiguously lead to the Onsager-Machlup action.
From a theoretical perspective, it will be interesting to explore
the relation between these two approaches to the stochastic action beyond the 
single-path limit of vanishing tube radius.
More explicitly, as formulated in Eq.~\eqref{eq:path_integral_limit},
 one can think of the finite-radius sojourn probability $P_R^{\varphi}(t_f)$ 
for a tube of radius $R$
around a reference path $\varphi$ as a path integral over 
an indicator function \cite{majumdar_persistence_1999,bray_persistence_2013}.
It will be interesting to see whether theoretical results like the 
 finite-radius sojourn probability \cite{kappler_stochastic_2020}
can also be calculated directly as path-integral averages within the
time-slicing path-integral formalism.
 }

\textcolor{black}{To summarize again,
with Eq.~\eqref{eq:LagrangianDifference} we provide an 
 intuitive and experimentally useful relation between observable exit rates and the stochastic
action Lagrangian.
We use this relation to show experimentally
 that the Onsager-Machlup Lagrangian characterizes
physical ratios of path probabilities.
More generally,
our work demonstrates that the asymptotic sojourn probability
provides a direct and experimentally accessible characterization of path properties in stochastic
dynamics.
We believe that both the theoretical
and experimental study of this observable 
\textcolor{black}{will be valuable for}
relating theoretical single-trajectory results to measurement in 
 stochastic dynamics.
}

\acknowledgements{
We thank Prof.~Mike Cates, Dr.~Yongjoo Baek, Dr.~Jules Guioth, Dr.~Rob Jack, 
and Dr.~Patrick Pietzonka for stimulating discussions.
In particular, we thank Dr.~Yongjoo Baek for
 pointing out that the results of Ref.~\cite{haken_generalized_1976,
wissel_manifolds_1979} can
be formulated elegantly via Eq.~\eqref{eq:DS_discrete2}.
Work was funded in part by the European Research Council under the EU's Horizon 2020 Program, Grant No.~740269, and by an Early Career Grant to R.A.~from the Isaac Newton Trust.
J.G.~and U.F.K.~were supported by the European Union's Horizon 2020 research and innovation program under European Training Network (ETN) Grant 674979-NANOTRANS. U.F.K.~acknowledges funding from 
an European Research Council Consolidator Grant (DesignerPores 647144).
}

\appendix

\renewcommand\thefigure{A\arabic{figure}}   
\counterwithin{figure}{section}

\section{Parametrizing the overdamped Langevin equation}
\label{app:parametrization}

We consider the overdamped one-dimensional It\^{o}-Langevin equation
\begin{equation}
\label{eq:LangevinEquation}
{\mathrm{d}X_t = D \beta F(X_t)\,\mathrm{d}t + \sqrt{2D}\, \mathrm{d}B_t}
\end{equation}
with $D$ the diffusivity,
$\beta^{-1} = k_{\mathrm{B}}T$ the thermal energy with $k_{\mathrm{B}}$
the Boltzmann constant and $T$ the absolute temperature,
$F(x) = - \nabla U$ an external force 
with a potential $U$, 
\textcolor{black}{
and $\mathrm{d}B_t$ the increment of the Wiener process.
Equation \eqref{eq:LangevinEquation} 
is identical to 
Eq.~\eqref{eq:ItoLangevin}
with $a(x) = D \beta F(x)$, $b = \sqrt{ 2 D}$.}
As we explain in the following, 
we parametrize Eq.~\eqref{eq:LangevinEquation} by locally calculating the first two 
Kramers-Moyal coefficients based on the experimental time series.
While this parameterization allows for a position-dependent diffusivity $D(x)$,
we will see below that for our experimental system the diffusivity is
well--approximated by a spatially constant diffusivity.
This in particular implies that, while we use the It\^{o} interpretation 
for
Eq.~\eqref{eq:LangevinEquation},
the choice of stochastic integral does
not lead to any ambiguity in our results, because for 
 constant diffusivity $D$ the It\^{o}- and Stratonovich interpretation of Eq.~\eqref{eq:LangevinEquation}
are equivalent \cite{kloeden_numerical_1992,gardiner_stochastic_2009}.

\begin{figure}[ht!]
\centering
	\includegraphics[width=\columnwidth]{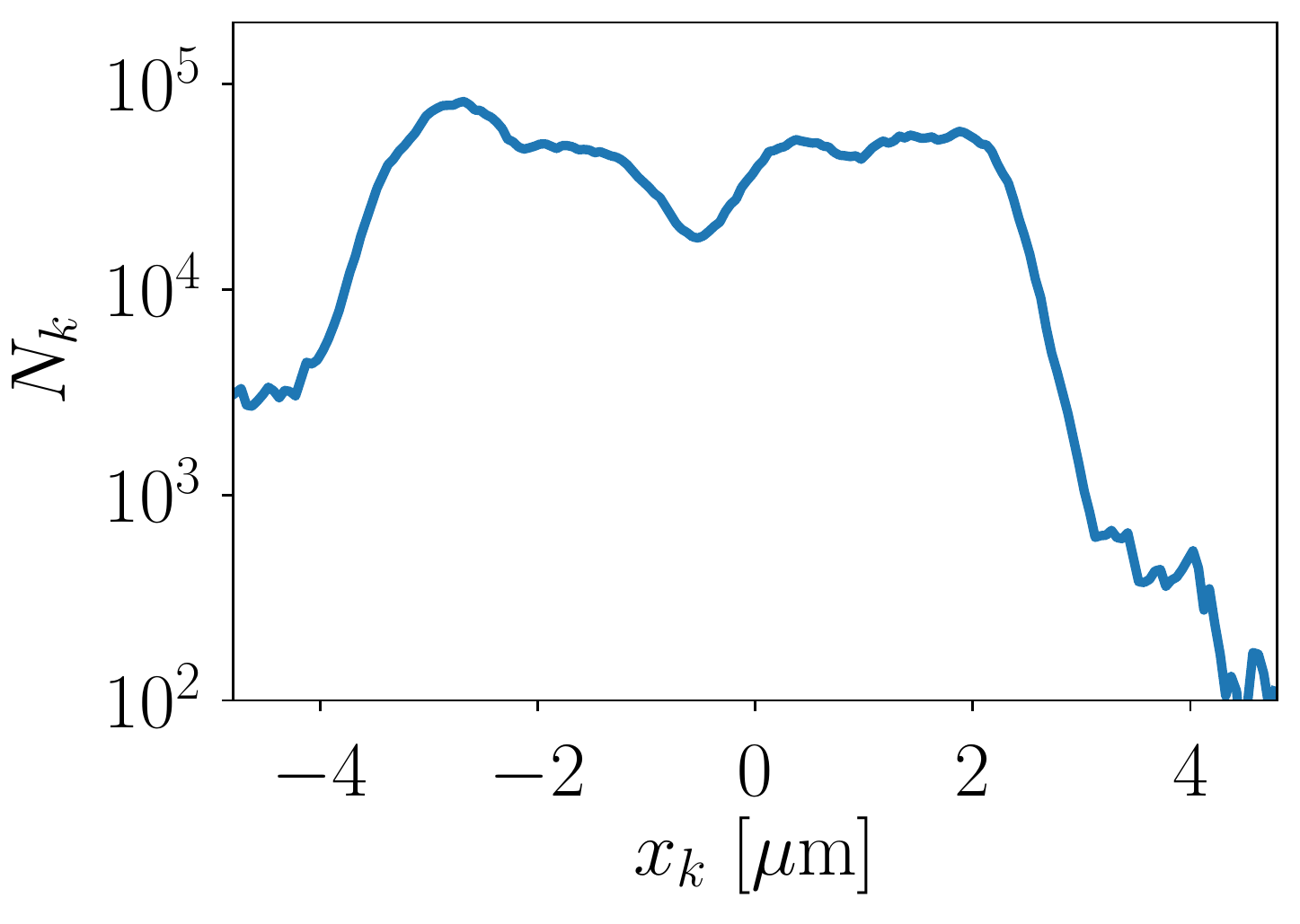}
\caption{
\label{fig:datapoints_per_bin}
\textit{Number of experimental data points per discretization bin.}
The solid line denotes the number of experimental data points per bin, as defined in 
Eq.~\eqref{eq:def_Nk}. 
The bin center ${x}_k$ is the center of the bin $B_k$,
defined in Eq.~\eqref{eq:def_Bk}.
}
\end{figure}

From 104 minutes of experimental measurements we obtain $N=230$ 
 uncorrelated discrete time series
\begin{equation}
	X_i(t_j) \equiv X_i (j \cdot \Delta t) \equiv X_{ij},
\end{equation}
where $i \in I = \{1,..,N\}$ labels the time series,
 and the maximal time index $j \in \{0,...,J_i\}$ for each time series
 depends on $i$, meaning the recorded time series are of variable length.
All time series have identical time step $\Delta t = 0.001\,$s, 
 lengths of time series range from 10 to 60 s.
We divide space into bins of width $\Delta x = 0.05\,\mu$m, with the $k$-th bin
\begin{equation}
\label{eq:def_Bk}
	B_k = \left[ \vphantom{e^{x^2}}\hat{x}_L + k\cdot \Delta x, ~\hat{x}_L + (k+1)\cdot \Delta x\right),
\end{equation}
where for the left boundary $\hat{x}_L= -4.8\,\mu$m,
and $k \in \{0,...,K\}$ with $K = 192$, so that 
$\hat{x}_R \equiv \hat{x}_L + K \cdot \Delta \hat{x} = 4.8\,\mu$m.
The center of the $k$-th bin, denoted by ${x}_k$, is located at
${x}_k \equiv \hat{x}_L + (k+1/2) \cdot \Delta x$.
The positions $\hat{x}_L$, $\hat{x}_R$ are still well within the experimental microchannel, 
meaning that a colloid starting at $\hat{x}_L$, $\hat{x}_R$ is very 
unlikely to leave the tube within one second.
For every bin $B_k$ we  create a list of all the experimentally recorded tuples $(i,j)$ such that 
$X_{ij} \in B_k$, i.e.~we for every $k$ construct the set
\begin{equation}
	\mathcal{B}_k = \left\{\,(i,j)\,\mid\,X_{ij} \in B_k\,\right\}.
\end{equation}
We denote the total number of data points in bin $B_k$ by 
\begin{equation}
\label{eq:def_Nk}
	N_k \equiv \left| \left\{ ~X_{ij} \in B_k~\right\} \right| \equiv |\mathcal{B}_k|,
\end{equation}
and show a plot of $N_k$ as a function of the bin center $x_k$
in Fig.~\ref{fig:datapoints_per_bin}.

To parametrize the overdamped Langevin Eq.~\eqref{eq:LangevinEquation},
we locally estimate both the diffusivity and the force  via  
discretized Kramers-Moyal coefficients \cite{gardiner_stochastic_2009}.
At the bin centered around ${x}_k$  we obtain
\begin{align}
\label{eq:diffusivity_formula}
	D({x}_k) &= \frac{1}{2 \cdot N_k \cdot \Delta t^*}\left[ \langle \Delta X^2(\Delta t^*) \rangle_k - \langle \Delta X(\Delta t^*)\rangle_k^2/N \right],\\
\label{eq:potential_formula}
	\beta F({x}_k) &= \frac{\langle \Delta X(\Delta t^*) \rangle_k}{D({x}_k)\cdot N_i \cdot\Delta t^*},
\end{align}
where the symbol $\langle \bullet \rangle_k$ denotes the average over all $N_k$ experimental
 time series
which start in the bin $B_k$. 
In the evaluation of Eqs.~\eqref{eq:diffusivity_formula}, \eqref{eq:potential_formula},
we furthermore use the lagtime $\Delta t^* = 15 \Delta t = 0.015$\,s, a discussion of the
dependence of our results on lagtime is given further below.
From the force we calculate a potential as
\begin{align}
\label{eq:potential_int}
\beta U(x_k) &= -\int_{\hat{x}_L}^{{x}_k}\mathrm{d}x'\,\beta F(x'),
\end{align}
where we use the trapezoidal rule to perform the integral on the right-hand side numerically;
the result of this integration is furthermore smoothed using a Hann-window that
at each $x_k$ incorporates the 20 closest datapoints.
The smoothed potential is then interpolated using polynomial splines of degree 3;
this polynomial interpolation is used in evaluations of the stochastic action to calculate the force $F$ and
 its derivative $\partial_x F$.

\begin{figure*}[ht!]
\centering
	\includegraphics[width=\textwidth]{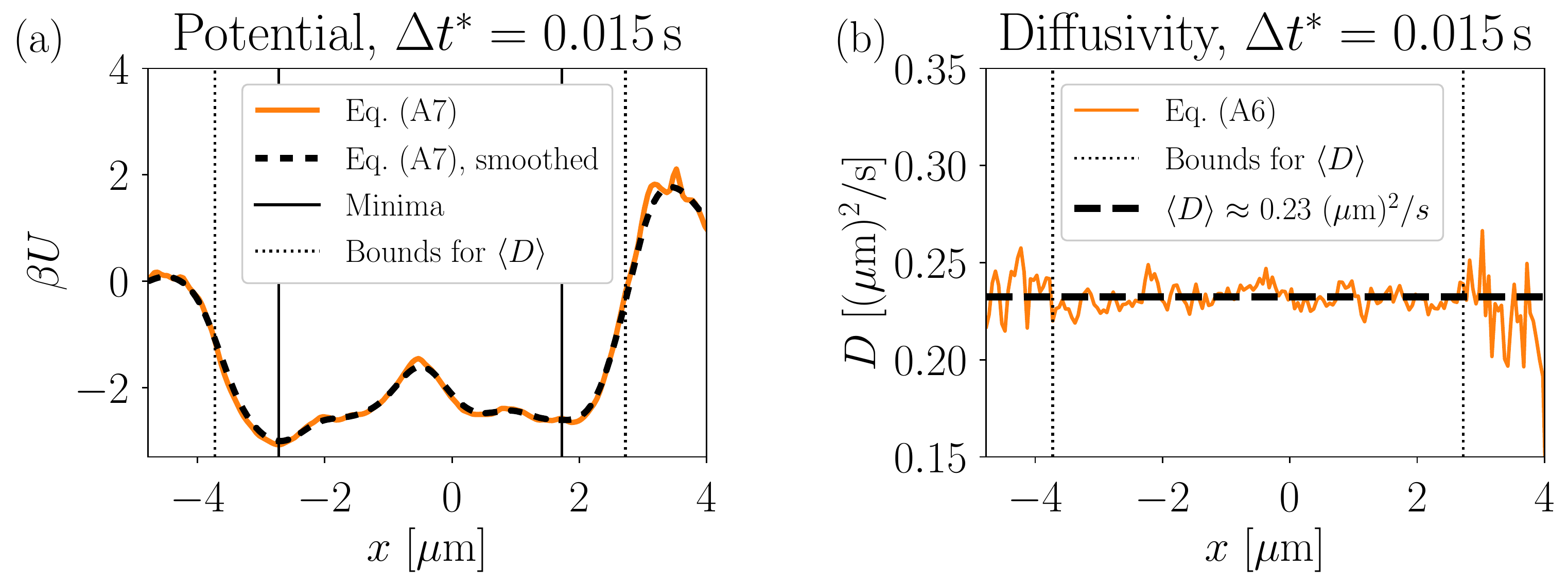}
\caption{
\label{fig:diffusivity_and_potential}
\textit{Local potential and diffusivity extracted from experimental time series.}
\textit{(a)}
The orange solid line depicts the potential energy as obtained from
 Eqs.~\eqref{eq:potential_formula}, \eqref{eq:potential_int}, for $\Delta t^* = 0.015$\,s.
The black dashed solid line is a smoothed version of the orange line, obtained via
 a Hann window average using 20 datapoints at each point $x_k$.
The vertical solid lines denote
local minima $\xminleft \approx -2.725\,\mu$m, 
$\xminright \approx 1.725\,\mu$m, of the smoothed potential energy.
The vertical dashed lines indicate the bounds 
of the interval $[\xminleft - 1\,\mu\mathrm{m},\xminright+ 1\,\mu\mathrm{m}]$ over which the average
diffusivity  $\langle D \rangle$ is calculated in subplot (b).
\textit{(b)} 
The orange line shows the local diffusivity as obtained from
Eq.~\eqref{eq:diffusivity_formula}.
The horizontal dashed line depicts the average over the diffusivity
 inside the interval $[\xminleft - 1\,\mu\mathrm{m},\xminright+ 1\,\mu\mathrm{m}]$, 
 as indicated by the two vertical lines, c.f.~subplot (a).
}
\end{figure*}

The diffusivity and potential energy profiles obtained 
from Eqs.~\eqref{eq:diffusivity_formula}, \eqref{eq:potential_formula} are shown
 in Fig.~\ref{fig:diffusivity_and_potential}.
The potential energy in subplot (a) shows two local minima at $\xminleft \approx -2.725\,\mu\mathrm{m}$, 
$\xminright \approx 1.725\,\mu\mathrm{m}$, separated by a barrier at $x \approx -0.5\,\mu\mathrm{m}$.
Note that in the main text a constant is added to the potential,
 such that the potential vanishes at $\xminleft$.
From Fig.~\ref{fig:diffusivity_and_potential} (b) we conclude that the
 diffusivity is almost independent of position within the interval
 $[\xminleft - 1\,\mu$m, $\xminright + 1\,\mu$m],
with an average value
 \begin{equation}
 \label{eq:mean_D}
 	\langle D \rangle \approx 0.232\,\frac{(\mu\mathrm{m})^2}{\mathrm{s}}.
\end{equation}

\begin{figure*}[ht!]
\centering
	\includegraphics[width=\textwidth]{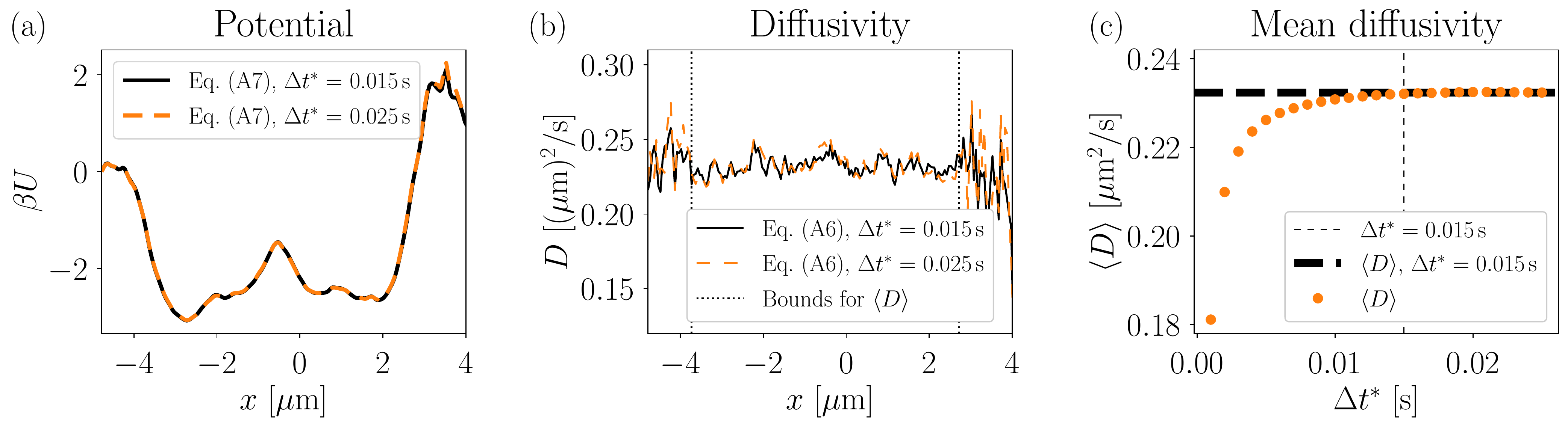}
\caption{
\label{fig:diffusivity_and_potential_lagtime_dependence}
\textit{Lagtime dependence of potential and diffusivity extracted from experiments.}
\textit{(a)}
The  potential energy as obtained from
 Eqs.~\eqref{eq:potential_formula}, \eqref{eq:potential_int}, 
 is shown for $\Delta t^* = 0.015$\,s
 as black solid line,
 and 
 for $\Delta t^* = 0.025$\,s
 as orange dashed line.
\textit{(b)}
 The diffusivity as obtained from
 Eq.~\eqref{eq:diffusivity_formula}
 is shown for $\Delta t^* = 0.015$\,s
 as black solid line,
 and 
 for $\Delta t^* = 0.025$\,s
 as orange dashed line.
The vertical dashed lines depict the boundary of the 
interval $[\xminleft - 1\,\mu\mathrm{m},\xminright+ 1\,\mu\mathrm{m}]$, 
where $\xminleft \approx -2.725\,\mu$m, $\xminright \approx 1.725\,\mu$m
are two local minima of the potential energy,
c.f.~Fig.~\ref{fig:diffusivity_and_potential}.
\textit{(c)}
Mean diffusivity $\langle D \rangle$, averaged over
$[\xminleft - 1\,\mu\mathrm{m},\xminright+ 1\,\mu\mathrm{m}]$,
 as function of lagtime $\Delta t^*$ used in Eq.~\eqref{eq:diffusivity_formula}.
 The horizontal thick dashed line denotes the value $\langle D\rangle$
 for $\Delta t^* = 0.015\,$s, as indicated by the vertical dashed line.
}
\end{figure*}

The dependence  of the inferred potential and diffusivity,
 Eqs.~\eqref{eq:diffusivity_formula}, \eqref{eq:potential_formula},
 on the lagtime $\Delta t^*$ is shown 
 in Fig.~\ref{fig:diffusivity_and_potential_lagtime_dependence}.
Subplots (a), (b) show that both the potential
and the diffusivity
 for the two lagtimes $\Delta t^* = 0.015$, $0.025\,$s, 
agree with each other.
Figure \ref{fig:diffusivity_and_potential_lagtime_dependence} (c) shows the average diffusivity
 $\langle D \rangle$
 as a function of the lagtime $\Delta t^*$. 
For short lagtimes $\Delta t^* \lesssim 0.01\,$s, the mean diffusivity slightly depends on the lagtime
(note the scaling on the $y$-axis),
which we attribute to inaccuracies of the centroid algorithm which we use to estimate colloidal positions.
For lagtimes $\Delta t^* \gtrsim 0.01\,$s, the mean diffusivity is independent of the lagtime, which justifies
our choice $\Delta t^* = 0.015\,$s.

\textcolor{black}{
Using the Einstein relation, the friction coefficient $\gamma$ follows from Eq.~\eqref{eq:mean_D}
as
\begin{equation}
\label{eq:mean_gamma}
	\gamma = \frac{k_{\mathrm{B}}T}{\langle D \rangle} \approx 1.75\cdot 10^{-8}\,\frac{\mathrm{kg}}{\mathrm{s}},
\end{equation}
where $k_{\mathrm{B}}$ is the Boltzmann constant and $T = 294\,\mathrm{K}$ is the experimental
temperature.
}

\textcolor{black}{
Figures \ref{fig:diffusivity_and_potential}, 
\ref{fig:diffusivity_and_potential_lagtime_dependence} demonstrate
 that,
 on the millisecond time scale,
  the dynamics of the colloidal particle along the channel axis
is \textcolor{black}{approximately Markovian and} 
well-described by an overdamped Langevin equation with additive noise.
This in particular implies that
both hydrodynamic interactions with the channel walls, as well as temporal noise correlations,
  are irrelevant on the scales we consider. We 
 now briefly discuss that this is consistent with other experimental studies involving colloidal particles. 
 }

 \textcolor{black}{
For colloidal particles in bulk water, 
deviations from white-noise behavior of the thermal force in fluids have been reported and characterized by Franosch {\em et al.} \cite{franosch_resonances_2011}. 
While the hydrodynamic-memory timescale $\tau_{\mathrm{f}}$ is of 
the order of 1 $\mu$s \cite{franosch_resonances_2011,lukic_motion_2007}, 
hydrodynamic effects can be observed on much larger timescales because
hydrodynamic noise correlations decay with a 
power-law tail.
More explicitly, in Ref.~\cite{franosch_resonances_2011} it was observed that
 colored deviations from white noise start to become relevant
 in bulk water on timescales slightly below 
  1 ms for beads of size 2-3 $\mu\mathrm{m}$. 
 For smaller colloids  with diameter 500 nm,   as used in our present work, we expect
 the onset of colored noise, and hence of hydrodynamic effects, to be on the order of 0.1 ms.
We expect this timescale to decrease even more
  in strong confinement (as compared to the
 corresponding bulk value), so that
  hydrodynamic memory effects are irrelevant on 
  the millisecond timescale probed in our measurements.
In this context we furthermore note that Ref.~\cite{lukic_motion_2007} also shows that 
 the inertial timescale $\tau_\textrm{p}$ of our colloidal particle is below 1 ${\mu}$s, which explains why
 inertial effects can be neglected in the overdamped Langevin model we use for
  our recorded data on the millisecond timescale.
  }
 
 \textcolor{black}{
The diffusivity of a colloidal particle in 
a confining microchannel is well-characterized experimentally \cite{dettmer_anisotropic_2014}.
In particular, the diffusivity is approximately
 position-independent in the interior of the channel, i.e.~sufficiently far away from the channel ends. 
The region of the channel shown in Fig.~1 (c), 
which we use for our measurements, is well within the channel,  so that
 as can be seen by the almost position-independent diffusivity in Fig.~A.2 (b), 
the boundary effects described in Ref.~\cite{dettmer_anisotropic_2014} are irrelevant here.
}

\section{Extracting sojourn probabilities from experimental time series}
\label{sec:algorithm}

\textit{Algorithm.}
We now explain how we extract sojourn probabilities and exit rates
from experimental time series. 
We assume as given several uncorrelated 
 time series, 
 a reference path $\traj(t)$, and a radius $R$. 
 In essence, the algorithm we use concatenates randomly sampled
  short recorded trajectories.

\textcolor{black}{We assume that the dynamics is time-homogeneous, 
and that the time series are Markovian.
The former assumption holds in our experimental data because the external
force is time-independent.
We discuss the validity of the  latter assumption in
Appendix~\ref{app:parametrization}.
We furthermore assume
that the time series
are indexed as described in the beginning of Appendix~\ref{app:parametrization}.}

At the initial time $\tinitial$,
we choose an initial probability density inside the tube. 
In the discretization of space described in Appendix~\ref{app:parametrization}, 
this probability density is represented by a normalized histogram that is 
only nonzero in the approximately $2R /\Delta x$ bins which intersect
 with the tube at time $\tinitial$, which is 
 given by the interval $[\varphi(\tinitial) -R,\varphi(\tinitial) + R]$.
To estimate the sojourn probability for a short time interval $\RefillTime$, we proceed as follows.
\begin{enumerate}
\item From the histogram representing the initial condition,
 we draw $M$  sample bins (with replacement) $\{B_{k_1}, B_{k_2},..., B_{k_M}\}$;
 for the definition of a bin see Eq.~\eqref{eq:def_Bk}.
Each sample bin represents an initial condition
 for a sample trajectory starting inside the tube.
\item For each sample bin $B_{k_i}$, we draw one
of the $N_{k_i}$ measured data points inside this bin
(with replacement, and using a uniform distribution on the set of all measured
data points inside the bin), where $N_i$ is defined in Eq.~\eqref{eq:def_Nk}.
If the bin $B_{k_i}$ only partly intersects the tube interior,
 and the drawn data point
lies outside the tube, a new datapoint is drawn.
The drawn datapoint belongs to a recorded time series, 
and we assume that this time series extends at least until 
time $t_{k_i} + \RefillTime$ (this requirement can always be ensured by
 reducing the maximal index $J_i$  corresponding to the trajectory $X_i$, 
and removing trajectories $X_i$ that are shorter than $\RefillTime$).
\item We follow each of the $M$ randomly drawn  time series from step 2 for the duration $\RefillTime$,
and discard each trajectory as soon as it first leaves the tube.
The number of trajectories left in the tube at each time step, denoted by $M_j$,
 yields an estimate for the sojourn probability via
	$P_R^{\traj}(j \cdot \Delta t) \equiv P_j \equiv  M_j/M$,
subject to the given initial condition,
and for a duration $\RefillTime$.
\item By creating a histogram from the final positions of those trajectories that stay 
inside the tube until time $\RefillTime$, 
a new initial distribution is obtained, and the algorithm can be repeated from step 1 for another time interval
$\RefillTime$.
\end{enumerate}
Figure \ref{fig:illustration} illustrates the algorithm for an initial distribution $P(x) = \delta (x -\varphi(\tinitial))$, 
$\RefillTime = 0.25\,\mathrm{s}$, $M=3$
(to obtain a reliable estimate for the sojourn probability, of course much larger values for $M$ need to be used).
For the analysis of the experimental data we use $\Delta x=0.05\,\mu$m, 
$\RefillTime = 0.25\,$s; 
at the end of the present section we show that results of this algorithm are independent 
of our particular choices for $\Delta x$ and  $\RefillTime$.

From the discrete time series $P_j$ for the sojourn probability, the exit rate $\aexit^{\traj}$
 is obtained by
discretizing
\begin{equation}
	\aexit^{\traj} = -\frac{\dot{P}_R^{\traj}(t)}{P_R^{\traj}(t)}.
\end{equation}

For the first iteration of steps 1-3 of the algorithm outline above, we choose $M = 10^5$
and as initial condition a smeared-out delta peak at the tube center, consisting of
a uniform distribution on the 3 bins closest to the tube center.
For each subsequent iteration of steps 1-3,
we estimate the number of trajectories $M$, based on the recent trend of the exit rate.
More explicitly, assuming we are at the $k$-th repetition of steps 1-3 (where $k > 1$),
we fit a linear function 
\begin{equation}
\label{eq:aexit_fit}
	\alpha_{\mathrm{fit}}(t) = a \cdot \left( t - (\tinitial + k \cdot \RefillTime)\right)+ b,
\end{equation}
to the exit rate in the time interval $[\tinitial + k \cdot \RefillTime-\Delta t_{\mathrm{fit}}, \tinitial + k \cdot \RefillTime]$, 
where $\Delta t_{\mathrm{fit}} = \min\{\,0.4\,$s, $\RefillTime\}$.
Using the fitted Eq.~\eqref{eq:aexit_fit}, we estimate the number $M$ such that the expected number
of trajectories inside the tube at the final time of the $k$-th iteration step is approximately $N_{\mathrm{final}}$,
which yields
\begin{align}
	N_{\mathrm{final}} &= M ~\exp\left[ - \int_{\tinitial + k \cdot \RefillTime}^{\tinitial + (k+1) \cdot \RefillTime} \mathrm{d}s ~\alpha_{\mathrm{fit}}(s) \right],
	\\
	\Longleftrightarrow \qquad
	M &= N_{\mathrm{final}} ~ \exp\left[ a \frac{\RefillTime^2}{2}   + b \,\RefillTime  \right].
	\label{eq:Nfinal_equation}
\end{align}
Unless noted otherwise, we use $N_{\mathrm{final}} = 10^5$
for all exit rates shown in this paper;
we demonstrate further below that our results are independent of the precise value used
for $N_{\mathrm{final}}$ (as long as it is sufficiently large).
For the minimization leading to the most probable path we also use smaller values
for $N_{\mathrm{final}}$, as described in detail in Appendix~\ref{app:instanton_from_data}.

\textcolor{black}{
\textit{Relative path likelihood for a pairs of paths.}
To infer the ratio of path probabilities for a pair of paths $\traj$, $\trajTwo$,
we use the algorithm described just above to
measure the exit rate for finite tube radius $R = 0.5$, 0.55, 0.6, 0.65, 0.7, 0.75, 0.8\,$\mu\mathrm{m}$.
Subsequently we extrapolate the corresponding finite-radius exit-rate difference 
\begin{equation}
\label{eq:exit_rate_difference_app}
\Delta \alpha_{R} (t) \equiv \alpha_R^{\varphi}(t) - \alpha_{R}^{\trajTwo}(t),
\end{equation}
to the limit $R\rightarrow 0$, as described in the main text.
For Fig.~\ref{fig:extrapolated_rate} in the main text, the path $\traj$ is parametrized as
\begin{equation}
\label{eq:ref_path_app}
	\varphi(t) = \frac{x_f-x_i}{2\arctan(\kappa/2)} \arctan\left[ \frac{\kappa}{t_f} \left( t -  t_f/2\right)\right] 
	 		+ \frac{x_f+x_i}{2},
\end{equation}
where $t = [0,t_f]$ with $ t_f =  20$\,s,
and where $x_i \equiv \xminleft \approx -2.725\,\mu$m, 
$x_f \equiv \xminright = 1.725\,\mu$m 
are two minima of the potential energy.
The path Eq.~\eqref{eq:ref_path_app}
describes a barrier crossing starting 
 at time $t_i=0$ at the left minimum and arriving at 
the right minimum at time $t_f = 20\,$s, with the parameter $\kappa$ controlling the maximal
path velocity during barrier crossing;
for the results shown in Fig.~\ref{fig:extrapolated_rate}, we use $\kappa = 5$.
For $\trajTwo$, we consider a path that rests at the right minimum, 
$\trajTwo(t) \equiv \xminright$, see Fig.~\ref{fig:extrapolated_rate} (a) for an illustration.
As we demonstrate further below, pairs in which both paths are time-dependent can also be considered;
the advantage of considering one constant path is that then all time-dependence
in the exit rate can be attributed to the non-constant path.
While in principle arbitrary paths can be considered, any path $\traj$ should of course move
so slowly that given the experimental time resolution $\Delta t$ of the data, the exit
rate from a tube of radius $R$ can be reliably inferred, i.e.~$\dot{\traj}\Delta t \ll R$.
}

\begin{figure*}[ht]
\centering
	\includegraphics[width=.9\textwidth]{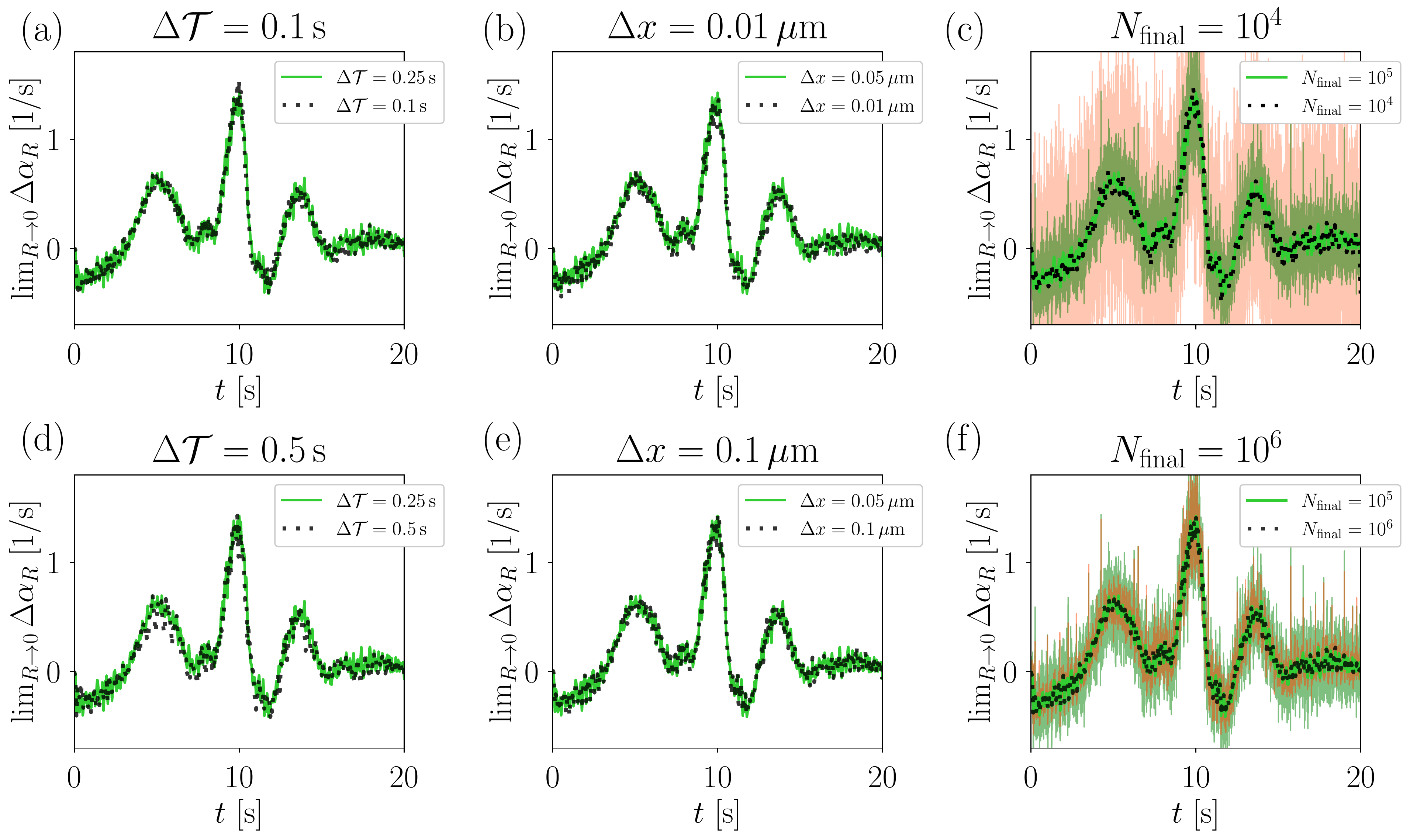}
\caption{
\label{fig:algorithm_invariance_refill}
\textcolor{black}{
\textit{Invariance of extrapolated exit-rate differences under variation of analysis parameters.}
All subplots feature a replot of the extrapolated exit rate from Fig.~\ref{fig:extrapolated_rate} (c)
of the main text (green solid line),
which is obtained using the algorithm from Appendix~\ref{sec:algorithm}
with $\RefillTime = 0.25\,$s, $\Delta x = 0.05\,\mu$m, $N_{\mathrm{final}} = 10^5$, and
the pair of reference paths shown in Fig.~\ref{fig:extrapolated_rate} (a) of the main text.
Subplots (a), (c) compare the extrapolated exit rate for $\RefillTime = 0.25\,$s (green solid line)
to results obtained using (a) $\RefillTime = 0.1\,$s and (d) $\RefillTime = 0.5\,$s (black dotted line),
with all other parameters for both analyses identical.
Subplots (b), (d) show both the extrapolated exit rate for $\Delta x = 0.05\,\mu$m (green solid line)
and the corresponding results extracted using (a) $\Delta x = 0.01\,\mu$m and (d) $\Delta x = 0.1\,\mu$m (black dotted line),
with all other parameters identical.
Subplots (c), (f) compare extrapolated exit-rate differences obtained for (c) $N_{\mathrm{final}} = 10^4$ and (d) $N_{\mathrm{final}} = 10^6$ to the results from
 Fig.~\ref{fig:extrapolated_rate} (c) of the main text, which use $N_{\mathrm{final}} = 10^5$.
For the latter data, the running average is plotted as thick green solid line,  the corresponding 
 unsmoothed time series
is shown as thin green solid line (and looks like a shaded area because of the short-timescale fluctuations
in the time series).
The running average corresponding to $N_{\mathrm{final}} = 10^4$, $10^6$ are shown as black dotted line,
the corresponding full unsmoothed time series is plotted as thin red line.
Except for the unsmoothed data in (c), (f), 
all shown data is smoothed using a running Hann-window average
with window width $0.1\,$s.
}
}
\end{figure*}

\textcolor{black}{
\textit{Invariance of algorithm under variation of $\RefillTime$.}
For the results shown in the main text we use 
$\RefillTime = 0.25\,$s.
To demonstrate that exit rates obtained using the algorithm described above
are independent of this particular choice of the parameter $\RefillTime$,
we now consider 
the exit rate difference of
 the pair of paths used in Fig.~\ref{fig:extrapolated_rate} of the main text
 for two other values of the parameter $\RefillTime$.
In Fig.~\ref{fig:algorithm_invariance_refill} (a), (d), we compare extrapolated exit-rate differences obtained for 
$\RefillTime = 0.1$, 0.5$\,$s, to results obtained using $\RefillTime = 0.25\,$s.
All curves show excellent agreement, so that we conclude that our results are independent of $\RefillTime$.
}

\textcolor{black}{
\textit{Invariance of results under variation of bin width $\Delta x$.}
In the algorithm described in the present appendix, particle positions are binned repeatedly,
and new samples of trajectories starting from those bins are drawn. We now demonstrate that
our results are independent of the particular bin width used for all results in the main text,
 $\Delta x = 0.05\,\mu$m.
For this, we consider the pair of paths from Fig.~\ref{fig:extrapolated_rate} 
of the main text, and infer ratios of path probabilities from the experimental data using
the bin widths $\Delta x = 0.01\,\mu$m and $\Delta x = 0.1\,\mu$m, with all other parameters 
identical to the  $\Delta x = 0.05\,\mu$m scenario.
The resulting extrapolated exit-rate differences are shown in Fig.~\ref{fig:algorithm_invariance_refill} (b), (e),
where we observe that the result is indeed independent of the bin width.
Note that the bin width $\Delta x = 0.01\,\mu$m is in fact of the order of the measuring error 
for the particle position (the experimental accuracy is about $0.05\,\mu$m); according to 
Fig.~\ref{fig:algorithm_invariance_refill} (b), 
this additional error does not influence the inferred extrapolated exit rate.
On the other hand, for $\Delta x = 0.1\,\mu$m the bin width is 10\% of the smallest tube diameter
considered, $2R = 1\,\mu$m; according to Fig.~\ref{fig:algorithm_invariance_refill} (e) 
this binning is still accurate enough to 
infer the extrapolated exit rate.
}

\textcolor{black}{
\textit{Invariance of results under variation of $N_{\mathrm{final}}$.}
For the parameter $N_{\mathrm{final}}$,
which via 
Eq.~\eqref{eq:Nfinal_equation} determines the number
 of trajectorial samples we consider in each iteration step to measure the exit rate,
we use $N_{\mathrm{final}} = 10^5$ in this work unless noted otherwise.
To show that our results do not depend on this particular choice, 
we in Fig.~\ref{fig:algorithm_invariance_refill} (c), (f)
show extrapolated exit-rate differences based on (c) $N_{\mathrm{final}} = 10^4$
and (f) $N_{\mathrm{final}} = 10^6$, with all other parameters identical to the
$N_{\mathrm{final}} = 10^5$ scenario.
As expected, we observe that for smaller $N_{\mathrm{final}}$, the fluctuations around the
running average of the extrapolated exit-rate difference are increased.
The running averages themselves agree very well for all values of $N_{\mathrm{final}}$
considered, so that we conclude that our results are independent of this parameter.
}

\begin{figure*}[ht]
\centering
	\includegraphics[width=.9\textwidth]{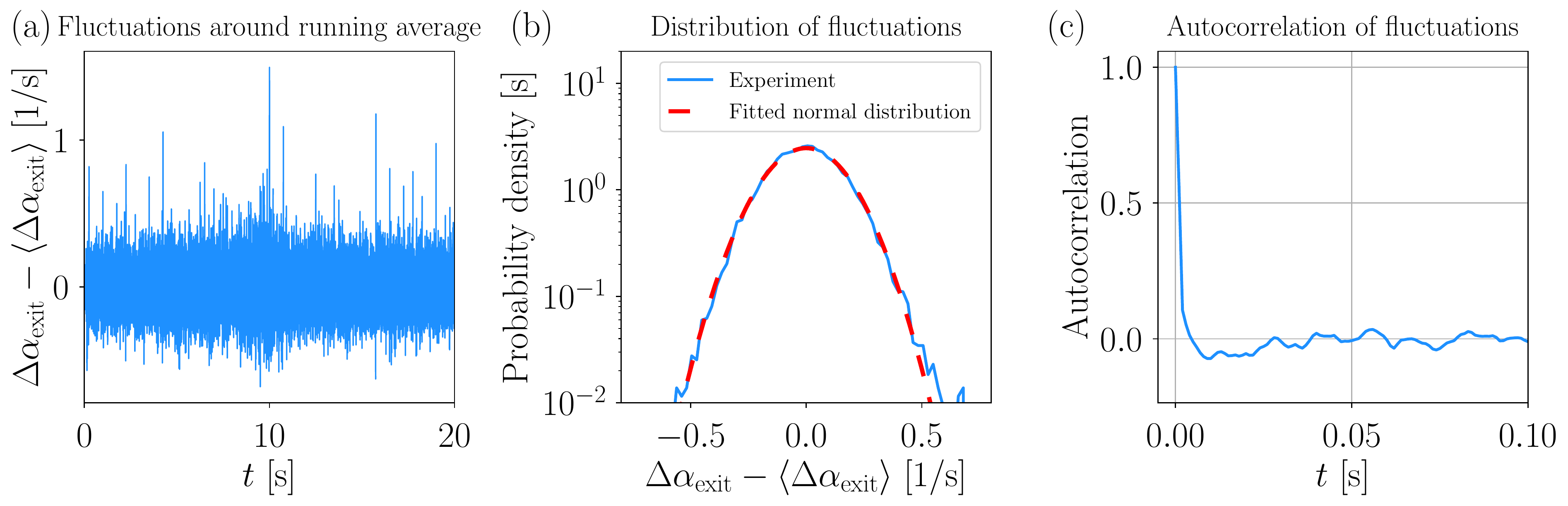}
\caption{
\label{fig:fluctuations}
\textcolor{black}{
\textit{Analysis of the noise eliminated by our smoothing procedure for extrapolated exit-rate differences.}
All data shown in this plot pertains to the smoothed extrapolated exit-rate difference considered
in Fig.~\ref{fig:extrapolated_rate} (c) of the main text, which is obtained using the algorithm
described in Appendix~\ref{sec:algorithm} with parameters $\Delta x = 0.05\,\mu$m,
$\RefillTime = 0.25\,$s, $N_{\mathrm{final}} = 10^5$.
Subplot (a) shows the deviation of the extrapolated exit-rate difference from its running average;
the latter is obtained from the former via a moving average with Hann-window of width $0.1\,$s.
Subplot (b) shows the distribution of the time series from subplot (a), 
together with a normalized fit of a Gaussian with zero mean;
the fitting interval is $[-0.5\,\mathrm{/s}, 0.5\,\mathrm{/s}]$,
the resulting variance is $\sigma = 0.16\,/$s.
Subplot (c) shows the normalized autocorrelation of the time series
shown in subplot (a) as a function of time.
}
}
\end{figure*}

\textcolor{black}{
\textit{Analysis of the noise eliminated by temporal averaging (smoothing).}
As can be seen in Figs.~\ref{fig:algorithm_invariance_refill} (c), (f), the 
experimental extrapolated exit-rate difference fluctuates significantly
around its running average.
We now show that these fluctuations are approximately normally distributed
and correlated only on very short timescales, which
indicates that they are approximately described by Gaussian white noise.
This justifies that we average over these fast fluctuations, which originate
from the fact that only a finite amount of experimental data is available.
For our analysis we consider the data used for Fig.~\ref{fig:extrapolated_rate} (c) of the main text,
which is based on analysis parameters $\Delta x = 0.05\,\mu$m, 
$\RefillTime = 0.25\,$s, and $N_{\mathrm{final}} = 10^5$,
and a moving Hann-window average with width $0.1\,$s.
For the discussion in this paragraph, we denote the unsmoothed extrapolated exit-rate difference 
by $\Delta \alpha_{\mathrm{exit}}$, and its running average  by
$\langle \Delta \alpha_{\mathrm{exit}}\rangle$ (this average corresponds to the
 green line shown in Fig.~\ref{fig:extrapolated_rate} (c) in the main text).
In Fig.~\ref{fig:fluctuations} (a) we show the fluctuations 
of $\Delta \alpha_{\mathrm{exit}}$ around its running average
as a function of time.
Surprisingly, the fluctuations are almost independent of time, 
and only slightly larger at around $t = 10\,$s, when the extrapolated exit-rate difference
is maximal, c.f.~Fig.~\ref{fig:extrapolated_rate} (c).
We note that there are discrete peaks, which occur at multiples of $\RefillTime$.
These originate from the instantaneous binning of the trajectories at these times
performed by our algorithm.
This binning slightly perturbs the instantaneous probability distribution inside the tube;
 the following relaxation of this perturbation, which happens on a very short time scale,
 leads to the observed peaks in the exit rate. 
In Fig.~\ref{fig:fluctuations} (b),
 we plot the
 distribution of the fluctuations from Fig.~\ref{fig:fluctuations} (a).
The distribution is approximately Gaussian, as shown by the included 
fit of a Gaussian distribution to the region $[-0.5\mathrm{/s},0.5\mathrm{/s}]$.
This interval contains 99.4\% of the fluctuations, 
which shows that the peaks observed at multiples of $\RefillTime$ in subplot (a) are statistically insignificant.
Figure \ref{fig:fluctuations} (c) displays the normalized autocorrelation function calculated from
the time series that shown in Fig.~\ref{fig:fluctuations} (a).
We observe that the autocorrelation decays on a time scale comparable to the
timestep of our data, and is completely uncorrelated for most of the 
duration of our averaging-window width $0.1\,$s.
\textcolor{black}{The short correlation time observed in the figure furthermore
 corroborates our assumption 
that the experimental time series is approximately Markovian, c.f.~Appendix~\ref{app:parametrization}.}
In conclusion, Fig.~\ref{fig:fluctuations} shows that the noise we eliminate by the running average
is approximately described by a stationary stochastic process with Gaussian steady-state
distribution and quickly decaying autocorrelation.
Our smoothing procedure thus basically eliminates Gaussian white noise, 
and thus does not introduce any spurious effects or bias
into the recorded data.
}

 \begin{figure}[ht]
\centering
	\includegraphics[width=.8\columnwidth]{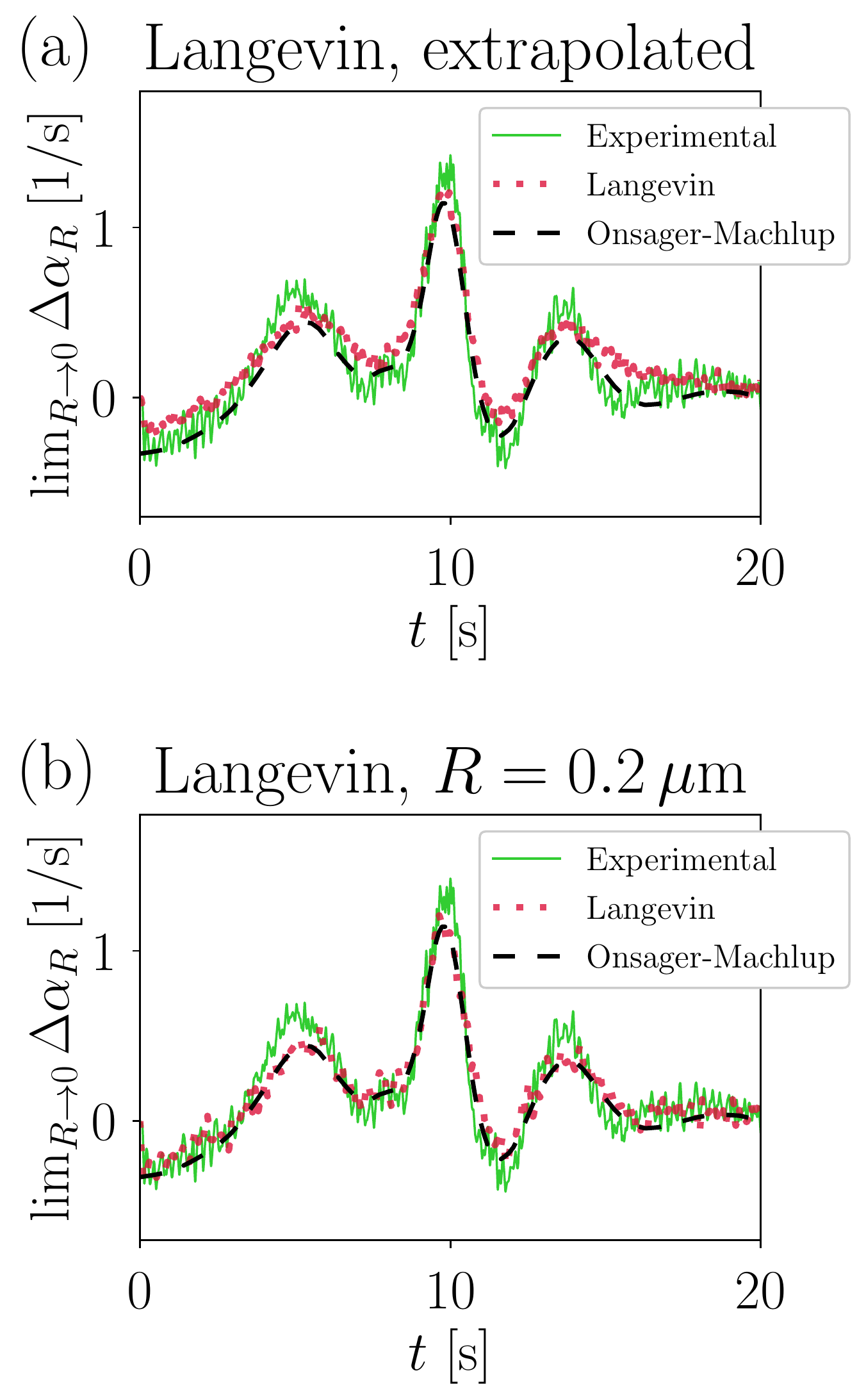}
\caption{
\label{fig:langevin}
\textcolor{black}{
\textit{Extrapolated and finite-radius exit-rate differences from Langevin simulations.}
This plot compares the extrapolated exit-rate difference
obtained directly from the experimental data
to results from Langevin simulations.
Subplot (a) shows a numerical extrapolated exit rate, obtained using a
 variation of the algorithm described in Appendix~\ref{sec:algorithm}, 
where sample trajectories are generated using Langevin simulations 
and no spatial binning is performed at multiples of $\RefillTime = 0.25\,$s, see Appendix~\ref{sec:algorithm} for details.
Subplot (b) shows the exit rate measured from Langevin simulations at finite radius $R = 0.2\,\mu$m.
All data is smoothed using a Hann-window average,
for the numerical data in subplot (b) we use
 an averaging window of width $0.2\,$s,
for all other shown data the averaging window has width $0.1\,$s.
}
}
\end{figure}

\textcolor{black}{
\textit{Extrapolated and finite-radius exit rates from Langevin simulations.}
In Fig.~\ref{fig:algorithm_invariance_refill} (b), (e), we show that 
the results of our algorithm are independent of the particular choice of the binning parameter
$\Delta x$.
As discussed before, 
the smallest feasible value for $\Delta x$ in our experimental analysis is determined by the accuracy
of our measurement apparatus. 
To demonstrate that even if there is no spatial averaging at all in our algorithm,
i.e.~in the limit $\Delta x \rightarrow 0$, the extrapolated exit rate is described by the OM Lagrangian,
we use a variation of our cloning algorithm to measure tubular exit rates from Langevin dynamics.
For this, we apply the algorithm to numerical simulations,
 but with two differences as compared to the analysis of experimental data.
First, we do not aggregate an ensemble of trajectories beforehand, but 
generate each sample trajectory on the fly
via an independent Langevin simulation in the potential energy landscape shown in 
Fig.~\ref{fig:extrapolated_rate} (c) of the main text
 using the constant diffusivity given by Eq.~\eqref{eq:mean_D},
and using a Euler-Maruyama integration scheme with a timestep $\Delta t_{\mathrm{num}} = 10^{-4}\,$s.
 Second, instead of binning positions after each iteration time 
  $\RefillTime = 0.25\,$s,
we directly sample initial conditions for the next iteration from the final positions of those
stochastic trajectories that have remained inside the tube;
this corresponds to a binning with vanishingly small bin width, $\Delta x \rightarrow 0$.
Using this algorithm, we measure the exit rate for 
the same values for $R$ and extrapolate to $R \rightarrow 0$
as for the experimental data.
The resulting extrapolated exit rate is compared to the experimental data
 in Fig.~\ref{fig:langevin} (a); we observe good agreement between the two, showing that
 our results are robust even in the limit $\Delta x \rightarrow 0$ (assuming that the stochastic
 process is described by an overdamped Langevin equation with additive noise).
As mentioned before,
the limiting factor for resolving the exit rate for small radius is both the temporal and spatial 
resolution; for our numerical simulations 
the temporal resolution is given by the integration timestep $\Delta t_{\mathrm{num}}$, and
the spatial resolution is basically the numerical integration error (which for the Euler-Maruyama 
algorithm and additive noise scales to leading order as $\Delta t_{\mathrm{num}}$).
Since the timestep in the simulations is a factor of ten smaller than the resolution of the experimental data,
using simulations we can measure the exit rate also at smaller radius.
Figure \ref{fig:langevin} (b) compares the exit-rate difference measured in Langevin
 simulations at the finite radius $R = 0.2\,\mu$m
to both the extrapolated experimental data and the OM Lagrangian; we observe that 
the finite-radius numerical result agrees well with the theoretical OM Lagrangian, indicating
that for the system parameters and paths considered here, for the radius $R = 0.2\,\mu$m
 the limit $R \rightarrow 0$ is almost realized.
 This agreement can be seen as direct numerical validation of the analytically calculated limit first
 obtained by
 Stratonovich \cite{stratonovich_probability_1971}.
}

\textcolor{black}{
\textit{Relative path likelihoods for several pairs of paths.}
}
 \begin{figure*}[ht]
\centering
	\includegraphics[width=\textwidth]{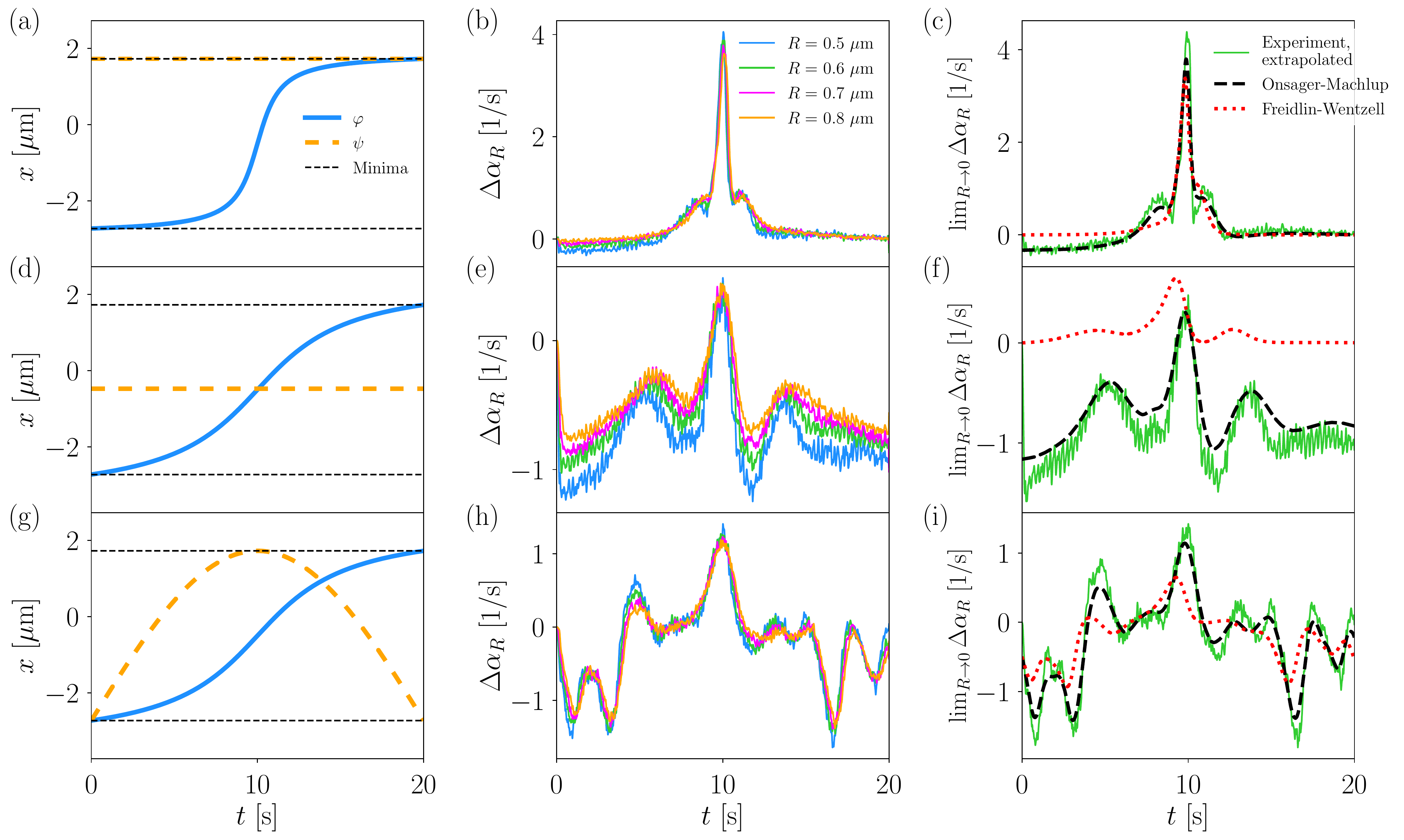}
\caption{
\label{fig:various_reference_paths}
\textcolor{black}{
\textit{(Extrapolated) exit-rate differences for various paths.}
Each subplot in the first column shows a pair of paths $\traj$ (blue solid line), 
$\trajTwo$ (orange dashed line),
as well as two local minima of the potential landscape from Fig.~\ref{fig:setup_and_potential} (c).
Each subplot in the second column depicts the measured 
 exit-rate differences Eq.~\eqref{eq:exit_rate_difference_app} 
 (colored solid lines),
as extracted directly from experimental time series for the paths from the first column
of each row, and for various values of the radius
 $R$, as indicated in the legend.
The third column shows the result of extrapolating the measured finite-radius exit-rate differences
for the paths from each row
to the limit $R \rightarrow 0$, as described in the main text.
Each subplot furthermore features the difference of FW and OM
Lagrangians evaluated along the corresponding path pairs.
Each exit rate shown in this figure
is smoothed using a Hann window 
of  width  $\Delta T_{\mathrm{smooth}} = 0.1\,$s. 
}
}
\end{figure*}
\textcolor{black}{
To demonstrate that relative path likelihoods inferred using our algorithm agree with the OM 
Lagrangian Eq.~\eqref{eq:OnsagerMachlup} for arbitrary pairs of
reference paths $\traj$, $\trajTwo$,
we now consider three more pairs of reference paths, 
illustrated in the first column of Fig.~\ref{fig:various_reference_paths}.
In each line, 
the second column shows the respective finite-radius exit-rate difference
Eq.~\eqref{eq:exit_rate_difference_app} obtained from experimental data.
The third column compares the resulting vanishing-radius extrapolations of the
measured exit-rate differences to the 
 Lagrangians Eqs.~\eqref{eq:OnsagerMachlup}, \eqref{eq:FreidlinWentzell}.
As can be seen, for all pairs of paths the experimental result agrees reasonably well
 with the OM Lagrangian, and shows clear disagreement with the FW Lagrangian. 
}

\textcolor{black}{
Note that also the OM Lagrangian sometimes does not agree perfectly with the experimental
data, which is most clearly observed in Fig.~\ref{fig:various_reference_paths} (f).
We believe that this is because even the smallest radii of our finite-radius tubes are not always
 small enough to perfectly capture the vanishing-radius limit.
Consequently, the agreement between experimental and theoretical results
 can presumably be improved upon by
 including measurements at smaller tube radius to the extrapolation.
This, however, requires a higher temporal and spatial resolution in the recorded time series,
which is ultimately limited by the experimental measurement apparatus.
}

\section{Calculating the most probable path from experimental data}
\label{app:instanton_from_data}

To extract the most probable path from experimental data,
we minimize the functional
\begin{equation}
\label{eq:functional_min_app}
\varphi^*_R \equiv \underset{\traj}{\mathrm{argmin}} \int_{t_i}^{t_f} \mathrm{d}t\,\alpha_{R}^{\traj}(t),
\end{equation}
for the finite values $R = 0.5$, 0.55, 0.6, 0.65, 0.7, 0.75, 0.8$\,\mu$m,
 and then extrapolate to $R = 0$.

For each $R$, the minimization in Eq.~\eqref{eq:functional_min_app}
is over all continuous paths with given
 endpoints $\traj(t_i) = x_i$, $\traj(t_f) = x_f$, so that 
 the minimization is over an
 infinite-dimensional space of functions.
To approximate this infinite-dimensional function space by
 a finite-dimensional space of dimension $N$, we parametrize  $\traj$ as
\begin{equation}
\label{eq:path_parametrization}
\traj(t) = x_i + \frac{t-t_i}{t_f-t_i} \left( x_f - x_i\right) + \sum_{n=1}^{N} \frac{a_n}{n^2} \sin\left( \pi n \frac{t-t_i}{t_f-t_i}\right).
\end{equation}
Note that for any given set of coefficients $(a_1,...,a_N)  \in \mathbb{R}^N$, Eq.~\eqref{eq:path_parametrization}
fulfills the boundary conditions $\traj(t_i) = x_i$, $\traj(t_f) = x_f$.
Employing this approximate parametrization, 
the minimization in Eq.~\eqref{eq:functional_min_app}
is, for given $R$, over $\mathbb{R}^N$.
Using our experimental data to evaluate the exit rate, 
we minimize the right-hand side of Eq.~\eqref{eq:functional_min_app},
  for  $N=20$ and $R = 0.5$, 0.55, 0.6, 0.65, 0.7, 0.75, 0.8$\,\mu$m 
  using a standard minimization algorithm \cite{hansen_cma-espycma_2019}.
For each evaluation of the sojourn
probability we employ the algorithm detailed in Appendix~\ref{sec:algorithm}.
Since the algorithm presented there
is based on stochastic sampling of recorded stochastic
trajectories, the sojourn probability obtained
from a single evaluation of our algorithm is 
 also stochastic. 
Using a larger value for $N_{\mathrm{final}}$ decreases the variance of the inferred
exit rate, but increases the computational time necessary to evaluate the exit rate for a given
reference path.

For computational efficiency, we proceed in several steps
to minimize Eq.~\eqref{eq:functional_min_app} for each given $R$.
\textcolor{black}{
First,
we perform four independent minimizations using
 $N_{\mathrm{final}}=2000$.
 For two of these minimizations we use as initial condition for the modes $a_n$, $n = 1,..., 20$,
 independent samples from a uniform distribution in $(-1,1)$.
 For the other two, we use as initial condition the minimum of the analytical FW and OM action 
 (each obtained as the lowest of 10 independent minimizations of the respective action using
  the potential energy and friction coefficient inferred from the measured data).
Each minimization has a starting variance
$\sigma_0 = 0.5$ for the minimization algorithm,
and we truncate the minimization after at most 2000 iterations of the algorithm
(during each iteration, the sojourn probability is evaluated 12 times); typically the minimization converges 
before that.
After these four minimizations,
the sojourn probability  for each of the four minima is evaluated again
using $N_{\mathrm{final}} = 10^5$, and the path with the largest sojourn probability is chosen
as $\varphi^*_R$.
}

Having obtained the most probable tube for several finite values of $R$, 
we subsequently  extrapolate
the corresponding modes $a_n(R)$,
 to $R=0$ by
fitting a function
$f_n(R) = A_n + R^2 B_n$
to the  finite-radius minimization results, 
and defining the corresponding expansion coefficients of the most probable path $\traj^*$ as
\begin{equation}
\label{eq:extrapolation_optimal_path}
a_n^* \equiv \lim_{R\rightarrow 0} f_n(R) = A_n.
\end{equation}

To minimize the OM and FW actions, obtained by integrating
 Eqs.~\eqref{eq:OnsagerMachlup}, \eqref{eq:FreidlinWentzell} along a path,
we also use the parametrization Eq.~\eqref{eq:path_parametrization},
with $N=40$;
the resulting instantons are shown in Fig.~\ref{fig:most_probable_path} (a).
Since the OM instanton agrees very well
with the experimental extrapolation, 
for which we use $N=20$,
we conclude that $N = 20$ modes are indeed sufficient to characterize the most probable 
path for the transition considered.

\section{Protocol for the comparison of FW and OM instantons}
\label{app:calculation_of_phase_diagram}

In  Fig.~\ref{fig:most_probable_path} (b) 
we show a contour plot of Eq.~\eqref{eq:FW_OM_diff} 
as a function of $T/T_0$ and $\TimeDiff = t_f - t_i$.
To obtain the figure, the actions corresponding to the OM and FW Lagrangians,
defined in Eqs.~\eqref{eq:OnsagerMachlup}, \eqref{eq:FreidlinWentzell},
are minimized using the experimental friction coefficient and force from Appendix~\ref{app:parametrization},
using a path parametrized via Eq.~\eqref{eq:path_parametrization} with $N=40$ modes.
Finding the most probable path for each action is then a nonlinear minimization
problem in $\mathbb{R}^N$.
\textcolor{black}{To carry out this minimization problem numerically
 we employ a cma-es algorithm \cite{hansen_cma-espycma_2019},
using the following protocol.
To ensure we find the global minimum for each parameter 
combination $( T/ T_0, \TimeDiff)$,
 we minimize each action in total 30 times. 
For every odd-numbered of these 30 minimizations,
we employ as initial condition for the minimizer a random initial condition where all the $a_n$, $n = 1,...,40$, 
are independent samples from a uniform distribution in $(-1,1)$;
 for every even-numbered of the 30 minimizations, we use the
 most probable OM/FW path from all the previous minimizations as initial condition for the respective other action.
 In all cases, the initial variance for the minimization algorithm is chosen as $\sigma_0 = 0.5$.
For each value of $(T/T_0,\TimeDiff)$, the
 respective action (FW/OM) is finally evaluated on all the results from the 30 minimizations,
and the path with the smallest action is used in Eq.~\eqref{eq:FW_OM_diff}.
For Fig.~\ref{fig:most_probable_path} (b), 
the resulting 2D array of data is
 subsequently smoothed using a Gaussian filter.}

\section{\textcolor{black}{Inferred instantons for various temperatures and total durations}}
\label{app:instantons_for_various_temperatures_and_durations}

\begin{figure*}[ht]
\centering
	\includegraphics[width=1\textwidth]{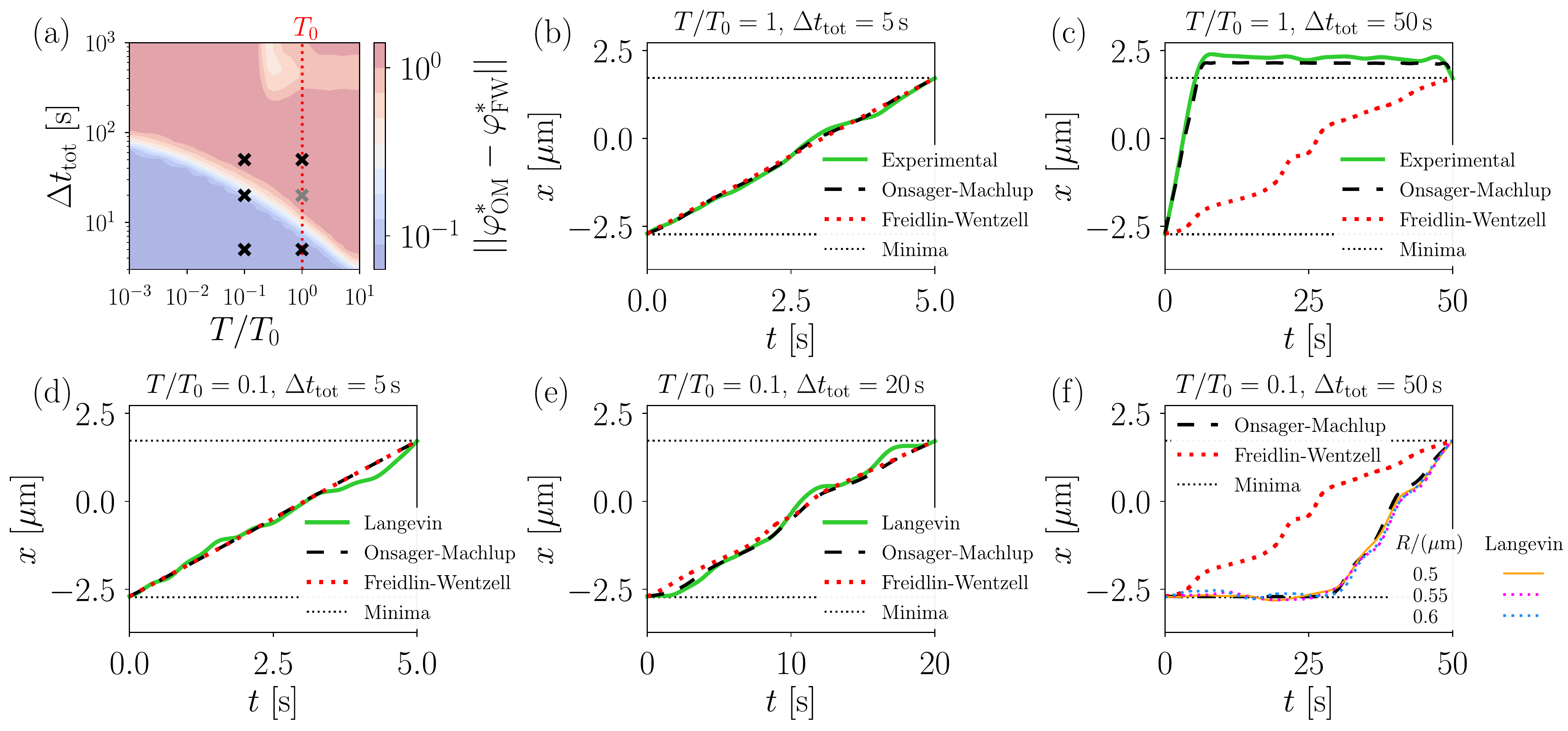}
\caption{
\label{fig:most_probable_paths_appendix}
\textcolor{black}{
\textit{Most probable paths for various
 temperatures $ T/ T_0$ and 
 total durations $\TimeDiff$.} 
Subplot (a) is a replot of Fig.~\ref{fig:most_probable_path} (b) of the main text,
and shows the mean difference between OM and FW instanton as a function 
of $ T/ T_0$ and $\TimeDiff$.
While the gray cross denotes the parameters  considered in 
Fig.~\ref{fig:most_probable_path} (a) of the main text,
the black crosses depict the parameters considered in subplots (b)-(f) of the present figure.
Subplots (b)-(f) show most probable paths 
  extracted from data using Eq.~\eqref{eq:extrapolation_optimal_path} (green solid line),
  together with the results of minimizing the integrated OM
 Lagrangian Eq.~\eqref{eq:OnsagerMachlup} (black dashed line), 
and minimizing the integrated 
FW Lagrangian Eq.~\eqref{eq:FreidlinWentzell} (red dotted line).
In each plot, the horizontal dashed lines denote the local minima
 $\xminleft \approx -2.725\,\mu$m, 
$\xminright \approx 1.725\,\mu$m,
 of the inferred potential, which are the start- and endpoints of the paths.
Subplots (b), (c), show most probable paths extracted 
 from the experimental data measured at $T_0 = 294\,$K.
Subplots (d), (e), (f), are obtained by analysis of an ensemble of numerical trajectories,
which is generated using Langevin simulations 
based on the friction coefficient and force profile measured at $T_0 = 294\,$K,
but with the random force scaled down to represent the temperature $T = T_0/10 = 29.4\,$K;
 see Appendix~\ref{app:instantons_for_various_temperatures_and_durations} for details.
 }
}
\end{figure*}

\textcolor{black}{
In Fig.~\ref{fig:most_probable_path} (b) of the main text
we discuss in which parameter regime FW and OM Lagrangians predict the same instanton.
In Fig.~\ref{fig:most_probable_path} (a) of the main text we compare
the functional minima for parameters $ T/ T_0 = 1$ and $\TimeDiff = 20\,$s 
 to the instanton extracted from experimental data, i.e.~we
 consider one particular point  of 
 Fig.~\ref{fig:most_probable_path} (b) of the main text
  (which is denoted by a black cross in the figure).
 We observe that, as expected from Figs.~\ref{fig:extrapolated_rate}
 and 
 \ref{fig:most_probable_path} (a) of the main text,
 the experimental result agrees with the 
 the OM instanton and disagrees with the FW instanton.
In the present appendix we consider several more points in the $( T/ T_0, \TimeDiff)$ plane,
namely $( T/ T_0, \TimeDiff /\mathrm{s}) = (1,5), (1,50), (0.1,5), (0.1,20), (0.1,50)$
as indicated in Fig.~\ref{fig:most_probable_paths_appendix} (a).
For each of these parameters, we extract the instanton from data and compare to the theoretical predictions 
from OM and FW Lagrangians.
}

\textcolor{black}{
For $ T/ T_0 = 1$ we 
apply the algorithm from Appendix~\ref{app:instanton_from_data}
to 
extract instantons
for $\TimeDiff = 5\,$s and $\TimeDiff = 50\,$s from the experimental data.}
\textcolor{black}{
For $ T/ T_0 = 1/10$, no experimental data is available.
Indeed, as we discuss in the main text, if our experimental system was cooled
 down to $T = T_0/10 = 29.4\,K$, the physics of the system would 
 be radically different from what we observe at room temperature,
  and presumably not described by the friction coefficient and force profile obtained
  at $T_0 = 294\,K$.
To obtain instantons based on trajectorial data also at $T/T_0 = 1/10$,
we generate an ensemble of Langevin trajectories, which we then analyze.
To generate the data, 
we use our inferred friction coefficient and force profile (inferred at temperature $T_0$),
 a diffusion coefficient
 that is rescaled to $T =  T_0/10$,
and run $10^4$ independent Langevin simulations
 of duration $2\,$s each, with a timestep $10^{-4}\,s$.
For each simulation, the initial condition is an independent sample of the
 uniform distribution in $[\xminleft-0.5\,\mu$m$,\xminright+0.5\,\mu$m$] 
 \approx [-3.225\,\mu$m$, 2.225\,\mu$m$]$.
We treat this ensemble of numerically generated trajectories
similar to the recorded experimental data, and
apply the algorithm from Appendix~\ref{app:instanton_from_data} to 
 find instantons for $\TimeDiff = 5$, $20$, $50\,$s.
Because for lower temperature the exit rate during barrier crossing is quite
large, we now consider a smaller refill time $\RefillTime = 0.025\,$s; this is unproblematic because the 
data is by construction perfectly Markovian.
For $\TimeDiff = 5$, $20\,$s we proceed exactly as with the experimental data, and extrapolate  inferred instantons for finite tube radius $R = 0.5$, $0.55$, $0.6$, $0.65$, $0.7$, $0.75$, $0.8\,\mu$m to vanishing
radius.
However, for $\TimeDiff = 50\,$s we observe that starting for $R \gtrsim 0.65\,\mu$m, the finite-radius
most probable tube starts to deviate a lot from the results for $R = 0.5$, $0.55$, $0.6\,\mu$m.
This indicates that for the larger radii we are too far away from the asymptotic vanishing-radius behavior  
of the exit rate to properly infer the limit $R \rightarrow 0$.
Therefore, instead of extrapolating to vanishing radius, for this case
we consider the finite-radius most probable tubes for $R = 0.5$, $0.55$, $0.6\,\mu$m.
}

\textcolor{black}{
In Fig.~\ref{fig:most_probable_paths_appendix} (b)-(f),
we compare the instantons obtained from experimental data (subplots (b), (c))
and Langevin data (subplots (d), (e), (f)) with the corresponding 
functional minima of the OM and FW action.
In all subplots, the experimental/Langevin instanton shows good
agreement with the OM instanton, and only agrees with the FW instanton
in the regime where OM and FW instanton agree with each other,
c.f.~Fig.~\ref{fig:most_probable_paths_appendix} (a).
}

\section{Continuum limit of the action for a realization of the Langevin equation}
\label{app:convergence}

\textcolor{black}{
We here discuss in more detail the continuum limit of the discretized action defined via
 Eqs.~\eqref{eq:continuum_limit_brownian2},
 \eqref{eq:DS_discrete2},
for the case that we evaluate the action on a realization of the Langevin Eq.~\eqref{eq:ItoLangevin},
which we write as $Y_t \equiv X_t$.
The results derived  in this appendix are well-known in the literature \cite{cameron_transformations_1944,girsanov_transforming_1960,
haken_generalized_1976,hunt_path_1981,oksendal_stochastic_2007},
and are included here for completeness.
}

\textcolor{black}{
Substituting Eq.~\eqref{eq:DS_discrete2} into the sum in Eq.~\eqref{eq:continuum_limit_brownian},
and expanding the square, the limit we are interested in can be written as
\begin{align}
&\SPI[X] \equiv\lim_{N\rightarrow \infty} \sum_{i=0}^{N-1}\Delta \SPI_i
\\&~~\nonumber
=\lim_{N\rightarrow \infty} \sum_{i=0}^{N-1} \frac{1}{b^2}\left[ 
\frac{1}{2} \frac{ \Delta X_i^2}{\Delta t} -  a(\bar{X}_i^{\dparam}) \Delta X_i 
+ \frac{1}{2} a(\bar{X}_i^{\dparam})^2 \Delta t\right]
\\ & \qquad- \lim_{N\rightarrow \infty} \sum_{i=0}^{N-1} 
 \frac{a'(\bar{X}^{\dparam}_i)}{2b^2} \left[ (1-2 \dparam) \Delta X_i^2 - b^2\Delta t\right]\,.
 \label{eq:continuum_limit_app_1}
\end{align}
As discussed in Eq.~\eqref{eq:diverging_term},
 the expectation value for the sum over the term $\Delta X_i^2/\Delta t$ in 
Eq.~\eqref{eq:continuum_limit_app_1}
 diverges in the continuum limit $N \rightarrow \infty$.
However, according to Eq.~\eqref{eq:diverging_term2} we have
\begin{equation}
\frac{1}{2b^2}
 \sum_{i=0}^{N-1}  
 \frac{ \Delta X_i^2}{\Delta t} = 
\frac{1}{2b^2}
 \sum_{i=0}^{N-1}  
 \frac{ \Delta \Bb_i^2}{\Delta t} + \mathcal{O}(N^{-1/2}),
\end{equation}
where $\Delta \Bb_i \equiv b \Delta B_i \equiv b \sqrt{\Delta t} \Delta W_i$ is the increment of the 
Wiener process, rescaled to the increment-variance at which it enters as random force
 in the Langevin equation, 
$\Bb_t \equiv b B_t$.
The diverging term in Eq.~\eqref{eq:continuum_limit_app_1} 
can therefore be removed by subtracting the discretized action $\SPI^{(0)}[\Bb]$ of 
 $\Bb$,
\begin{equation}
\sum_{i=0}^{N-1} \Delta \SPI_{i}^{(0)} \equiv\frac{1}{2b^2} \sum_{i=0}^{N-1} \Delta t\left( \frac{ \Delta \Bb_i}{\Delta t}\right)^2,
\end{equation}
 before taking the continuum limit $N \rightarrow \infty$. 
This means
 that instead of Eq.~\eqref{eq:continuum_limit_app_1} we consider the limit
\begin{align}
&\SPI[X] -\SPI^{(0)}[\Bb] 
\\ &\quad \equiv\lim_{N\rightarrow \infty}  \sum_{i=0}^{N-1}\left[ \Delta S_i (X_{i+1},X_i,\Delta t)
- \frac{1}{2b^2}\frac{ \Delta \Bb_i^2}{\Delta t}\right]
\label{eq:app_continuum_limit_three_sums_}
\\ &\quad = 
- \lim_{N\rightarrow \infty} \sum_{i=0}^{N-1} \frac{1}{b^2} a(\bar{X}_i^{\dparam}) \Delta X_i 
 \nonumber
\\ &\qquad 
+ \lim_{N\rightarrow \infty} \sum_{i=0}^{N-1} \frac{1}{2b^2} a(\bar{X}_i^{\dparam})^2 \Delta t
\label{eq:app_continuum_limit_three_sums}
\\ & \qquad- \lim_{N\rightarrow \infty} \sum_{i=0}^{N-1} 
 \frac{a'(\bar{X}^{\dparam}_i)}{2b^2} \left[ (1-2 \dparam) \Delta X_i^2 - b^2\Delta t\right].
 \nonumber
\end{align}
This amounts to considering
the limiting ratio
\begin{equation}
\label{eq:limiting_ratio_girsanov}
\lim_{N \rightarrow \infty} 
\frac{ P(X_N,t_N;...;X_1,t_1\mid X_0,t_0)}
{
 P^{\mathbb{W}}(\Bb_N,t_N;...;\Bb_1,t_1\mid \Bb_0,t_0)
}
\equiv 
e^{-(\SPI[X]-\SPI^{(0)}[\Bb])},
\end{equation}
where the $N$-point probability density in the numerator of Eq.~\eqref{eq:limiting_ratio_girsanov}
is with respect to the Langevin dynamics,
whereas the $N$-point probability density in the denominator
is with respect to the rescaled Wiener process $\Bb$, as indicated by the superscript $\mathbb{W}$.
For the two diverging sums in Eq.~\eqref{eq:app_continuum_limit_three_sums_}
to cancel,
the Langevin realization in the numerator of Eq.~\eqref{eq:limiting_ratio_girsanov} 
must correspond to the same random force realization as used
in the denominator;
 because  the expression 
 depends only on the increments of the Wiener process,
the initial value of the Wiener process, $\Bb_0$, is in fact irrelevant.
}

\textcolor{black}{
The limiting ratio Eq.~\eqref{eq:limiting_ratio_girsanov}
quantifies the probability ratio of a realization $X_t$ of the 
Langevin Eq.~\eqref{eq:ItoLangevin}
and the  {corresponding}
 realization of the Brownian motion $\Bb_t$.
The limit Eq.~\eqref{eq:app_continuum_limit_three_sums} 
therefore relates the probability distributions induced on the space of continuous paths
by the Langevin Eq.~\eqref{eq:ItoLangevin} to the probability distribution induced
on the same space  by the (rescaled) Wiener process.
This relation is described by
 the Girsanov formula,
and indeed we will see further below that the limit Eq.~\eqref{eq:limiting_ratio_girsanov}
is precisely the Radon-Nikodym
 derivative from the Girsanov theorem
\cite{cameron_transformations_1944,girsanov_transforming_1960,oksendal_stochastic_2007}.
Even more, the time-slicing approach employed here was used by Cameron
and Martin to derive an early variant of the Girsanov theorem \cite{cameron_transformations_1944}.
}

\textcolor{black}{
To calculate   $\SPI[X] - \SPI^{(0)}[\Bb]$, and to see explicitly that the result is in fact independent of $\dparam$,
we now consider the limit for each of the sums in Eq.~\eqref{eq:app_continuum_limit_three_sums} 
separately.
Conceptually, it is clear without calculation that the final result of the following calculation
 must not dependent on $\dparam$, as the sum over the terms Eq.~\eqref{eq:DS_discrete2} is
  asymptotically independent of $\dparam$.
}
 
 \textcolor{black}{
The first sum in Eq.~\eqref{eq:app_continuum_limit_three_sums}  is simply the definition
of the stochastic integral \cite{gardiner_stochastic_2009}, 
using the convention of evaluating the integrand at an intermediate point given by
$\dparam$, which we write as
\begin{equation}
\lim_{N\rightarrow \infty} \sum_{i=0}^{N-1} a(\bar{X}_i^{\dparam}) \Delta X_i \equiv \int_0^{t_f} 
\circled[0]{\footnotesize{$\dparam$}}\,
\mathrm{d}X_t\, a({X}_t),
\label{eq:app_first_term_0}
\end{equation}
where the symbol $\circled[0]{\footnotesize{$\dparam$}}$ indicates the convention for the stochastic integral.
For example, for $\dparam = 0$ we obtain the It\^{o} stochastic integral, whereas for $\dparam = 1/2$
we obtain the Stratonovich stochastic integral \cite{gardiner_stochastic_2009}.
For any value of $\dparam$, the stochastic integral along a solution of the Langevin equation
can be rewritten in terms of an It\^{o} integral
\cite{hunt_path_1981,gardiner_stochastic_2009}. 
This is achieved by Taylor expanding $a(\bar{X}_i^{\dparam})$
 around $X_i$, 
\begin{align}
\label{eq:app_first_term_1}
a(\bar{X}^{\dparam}_i = X_i + \dparam \Delta X_i) &= a(X_i) + \dparam \Delta X_i a'(X_i) + \mathcal{O}(\Delta t),
\end{align}
where we use that
 from Eq.~\eqref{eq:order_15_discretization} it follows that $\Delta X_i = \mathcal{O}(\Delta t^{1/2})$. 
Substituting Eq.~\eqref{eq:app_first_term_1} into the left-hand side of Eq.~\eqref{eq:app_first_term_0},
we obtain
\begin{align}
\nonumber
\lim_{N\rightarrow \infty} \sum_{i=0}^{N-1} a(\bar{X}_i^{\dparam}) \Delta X_i 
&
=
\lim_{N\rightarrow \infty} \sum_{i=0}^{N-1} a(X_i) \Delta X_i 
\\ &\quad 
\nonumber
+ 
\dparam \lim_{N\rightarrow \infty} \sum_{i=0}^{N-1} a'(X_i) \Delta X_i^2 
\\ &\quad
+
\lim_{N\rightarrow \infty} \sum_{i=0}^{N-1} \mathcal{O}(\Delta t) \Delta X_i.
\label{eq:app_first_term_2}
\end{align}
The first sum in Eq.~\eqref{eq:app_first_term_2} is the It\^{o} stochastic integral, which we denote by
\begin{equation}
\label{eq:app_first_term_3}
\lim_{N\rightarrow \infty} \sum_{i=0}^{N-1} a(X_i) \Delta X_i  \equiv \int_0^{t_f} \mathrm{d}X_t \,a(X_t).
\end{equation}
Substituting 
$\Delta X_i = b \Delta t^{1/2}\Delta W_i + \mathcal{O}(\Delta t)$,
which follows from Eq.~\eqref{eq:order_15_discretization},
into the second sum in Eq.~\eqref{eq:app_first_term_2}, we obtain 
\begin{align}
\lim_{N\rightarrow \infty} \sum_{i=0}^{N-1} a'(X_i) \Delta X_i^2 
&=
b^2 \lim_{N\rightarrow \infty} \sum_{i=0}^{N-1} a'(X_i) \Delta t \Delta W_i^2 
\\ & =
b^2 \int_0^{t_f} \mathrm{d}t\, a'(X_t).
\label{eq:app_first_term_4}
\end{align}
The last equality 
follows via exactly the same
argument as employed in proving the
 It\^{o} formula, see e.g.~Chapter 4 of Ref.~\cite{oksendal_stochastic_2007}.
As stated in Ref.~\cite{oksendal_stochastic_2007}, 
we note that because $\Delta {B}_i \equiv \sqrt{ \Delta t} \Delta W_i$
is the increment of the Wiener process,
 the limit Eq.~\eqref{eq:app_first_term_4} is often expressed
by the formula $\mathrm{d}{B}_t^2 = \mathrm{d}t$.
Because $ \mathcal{O}(\Delta t) \Delta X_i = \mathcal{O}(\Delta t^{3/2}) = \mathcal{O}(N^{-3/2})$, 
 the last sum in Eq.~\eqref{eq:app_first_term_2}
scales as $\mathcal{O}(N^{-1/2})$ and hence vanishes in the limit $N \rightarrow \infty$.
}

\textcolor{black}{
Combining Eqs.~\eqref{eq:app_first_term_0}, 
\eqref{eq:app_first_term_2}, 
\eqref{eq:app_first_term_3}, 
\eqref{eq:app_first_term_4}, 
we obtain
\begin{align}
\lim_{N\rightarrow \infty} \sum_{i=0}^{N-1} a(\bar{X}_i^{\dparam}) \Delta X_i &\equiv \int_0^{t_f} 
\circled[0]{\footnotesize{$\dparam$}}\,
\mathrm{d}X_t\, a({X}_t)
\\ & =
\int_0^{t_f} 
\mathrm{d}X_t\, a({X}_t)
\label{eq:app_first_term_ito_vs_alpha}
\\ & \qquad + \nonumber
b^2 \dparam
\int_0^{t_f} 
\mathrm{d}t\, a'({X}_t),
\end{align}
where the last equality is the standard formula that relates different definitions
of the stochastic integral along the solution of the Langevin equation
 to the It\^{o} convention \cite{hunt_path_1981,gardiner_stochastic_2009}.
 }

\textcolor{black}{
The second sum in Eq.~\eqref{eq:app_continuum_limit_three_sums} yields
\begin{align}
\label{eq:app_second_sum_result}
 \lim_{N\rightarrow \infty} \sum_{i=0}^{N-1} \frac{1}{2b^2} a(\bar{X}_i^{\dparam})^2 \Delta t
 &=
 \frac{1}{2b^2} \int_0^{t_f} \mathrm{d}t\,a({X}_t)^2.
\end{align}
That the result is independent of $\dparam$ follows by Taylor expanding $a(\bar{X}_i^{\dparam})$
around $X_i$, c.f.~Eq.~\eqref{eq:app_first_term_1}.
Similar to the third sum in Eq.~\eqref{eq:app_first_term_2}, 
 the left-hand side of Eq.~\eqref{eq:app_second_sum_result}
then depends on $\dparam$
only  via
terms that vanish in the continuum limit $N \rightarrow \infty$.
}

\textcolor{black}{
The continuum limit of the third sum in Eq.~\eqref{eq:app_continuum_limit_three_sums}
is given by
\begin{align}
\label{eq:convergence_eq}
\lim_{N\rightarrow \infty}
\sum_{i=0}^{N-1} \frac{a'(\bar{X}^{\dparam}_i)}{2b^2} &\left[ (2 \dparam-1) \Delta X_i^2 + b^2\Delta t\right] 
\\ &\qquad\qquad=\dparam \int_0^{t_f}\mathrm{d}t \,a'(X_t).
\nonumber
\end{align}
To obtain this, we first note that,
\begin{equation}
\label{eq:app_sum_0}
\lim_{N\rightarrow \infty}\sum_{i=0}^{N-1} a'(\bar{X}^{\dparam}_i) \Delta t = \int_0^{t_f}\mathrm{d}t\,a'(X_t),
\end{equation}
which follows from the same arguments as used in Eq.~\eqref{eq:app_second_sum_result}.
The other limiting sum in Eq.~\eqref{eq:convergence_eq} is given by
\begin{equation}
\label{eq:convergence_0}
\lim_{N \rightarrow \infty}\sum_{i=0}^{N-1} a'(\bar{X}^{\dparam}_i) \Delta X_i^2 = b^2 \int_0^{t_f} \mathrm{d}t \,a'(X_t),
\end{equation}
which follows from the same argument as used
 in Eq.~\eqref{eq:app_first_term_4} \cite{oksendal_stochastic_2007}.
Upon substituting the two limits Eqs.~\eqref{eq:app_sum_0}, \eqref{eq:convergence_0}
into the left-hand side of Eq.~\eqref{eq:convergence_eq}, the right-hand side 
of the equation follows.
}

\textcolor{black}{
Finally substituting Eqs.~\eqref{eq:app_first_term_0}, 
\eqref{eq:app_second_sum_result}, 
\eqref{eq:convergence_eq} into Eq.~\eqref{eq:app_continuum_limit_three_sums}, we obtain \cite{hunt_path_1981}
\begin{align}
\SPI[X] -\SPI^{(0)}[\Bb]  &=
-\frac{1}{b^2}
\int_0^{t_f} 
\circled[0]{\footnotesize{$\dparam$}}\,
\mathrm{d}X_t\, a({X}_t)
+ \frac{1}{2b^2} \int_0^{t_f} \mathrm{d}t\,a({X}_t)^2
\nonumber
\\ & \quad
+
\dparam \int_0^{t_f}\mathrm{d}t \,a'(X_t).
\label{eq:girsanov_in_alpha_discretization}
\end{align}
While the symbol $\dparam$ appears explicitly in this expression, the independence of
$S[X_t] -S_0[B_t]$ on $\dparam$ becomes apparent 
by transforming the stochastic integral in the equation to
the It\^{o} convention via Eq.~\eqref{eq:app_first_term_ito_vs_alpha}, which leads to 
\begin{align}
\label{eq:girsanov_in_ito_discretization}
\SPI[X] -\SPI^{(0)}[\Bb]  &=
-\frac{1}{b^2}\int_0^{t_f} 
\mathrm{d}X_t\, a({X}_t)
+ \frac{1}{2b^2} \int_0^{t_f} \mathrm{d}t\,a({X}_t)^2.
\end{align}
From this manifestly $\dparam$-independent expression it is also apparent that 
the limit Eq.~\eqref{eq:limiting_ratio_girsanov}
is the Radon-Nikodym
derivative of the Langevin trajectory $X$ with respect to its corresponding
noise realization $\Bb$, as described by the 
Girsanov formula \cite{cameron_transformations_1944,
girsanov_transforming_1960,
oksendal_stochastic_2007}.
}

\end{document}